\documentclass[trackchanges, twocolumn]{aastex6}
\usepackage{amsmath}
\usepackage{graphicx}
\usepackage{algorithmicx,algpseudocode}
\usepackage{multirow}
\usepackage{color}

\newcommand{\mcr}[1]{\multicolumn{1}{r}{#1}}
\newcommand{\vone}{\mathbf{1}}
\newcommand{\vb}{\mathbf{b}}
\newcommand{\vh}{\mathbf{h}}
\newcommand{\vk}{\mathbf{k}}
\newcommand{\valpha}{\boldsymbol{\alpha}}
\newcommand{\vbeta}{\boldsymbol{\beta}}
\newcommand{\vtheta}{\boldsymbol{\theta}}
\newcommand{\vgamma}{\boldsymbol{\gamma}}
\newcommand{\bSigma}{\boldsymbol{\Sigma}}

\newcommand{\vmu}{\boldsymbol{\mu}}
\newcommand{\vy}{\mathbf{y}}
\newcommand{\vrt}{\mathbf{r}}
\newcommand{\vzero}{\mathbf{0}}
\newcommand{\calGP}{\mathcal{GP}}
\newcommand{\vt}{\mathbf{t}}
\newcommand{\K}{\mathbf{K}}
\renewcommand{\H}{\mathbf{H}}
\newcommand{\I}{\mathbf{I}}

\newcommand{\calN}{\mathcal{N}}

\DeclareMathOperator*{\argmin}{arg\,min}
\DeclareMathOperator*{\argmax}{arg\,max}
\shorttitle{Semi-parametric period estimation}
\shortauthors{He, Yuan, Huang, Long \& Macri}

\begin{document}
\title{Period estimation for sparsely sampled quasi-periodic\\light curves applied to Miras}

\author{Shiyuan He\altaffilmark{1}, Wenlong Yuan\altaffilmark{2}, Jianhua Z. Huang\altaffilmark{1}, James Long\altaffilmark{1} \& Lucas M. Macri\altaffilmark{2,3}}

\altaffiltext{1}{Department of Statistics, Texas A\&M University, College Station, TX, USA}
\altaffiltext{2}{George P.~and Cynthia W.~Mitchell Institute for Fundamental Physics \& Astronomy, Department of Physics \& Astronomy,\\Texas A\&M University, College Station, TX, USA}
\altaffiltext{3}{Corresponding author, {\tt lmacri@tamu.edu}}

\begin{abstract}
We develop a nonlinear semi-parametric Gaussian process model to estimate periods of Miras with sparsely sampled light curves. The model uses a sinusoidal basis for the periodic variation and a Gaussian process for the stochastic changes. We use maximum likelihood to estimate the period and the parameters of the Gaussian process, while integrating out the effects of other nuisance parameters in the model with respect to a suitable prior distribution obtained from earlier studies. Since the likelihood is highly multimodal for period, we implement a hybrid method that applies the quasi-Newton algorithm for Gaussian process parameters and search the period/frequency parameter space over a dense grid.

A large-scale, high-fidelity simulation is conducted to mimic the sampling quality of Mira light curves obtained by the M33 Synoptic Stellar Survey. The simulated data set is publicly available and can serve as a testbed for future evaluation of different period estimation methods. The semi-parametric model outperforms an existing algorithm on this simulated test data set as measured by period recovery rate and quality of the resulting Period-Luminosity relations.
\end{abstract}


\section{Introduction}
The determination of reliable periods for variable stars has been an area of interest in astronomy for at least four centuries, since the discovery of the variability of Mira ($o$~Ceti) by Fabricius in 1596 and the first attempts to determine its period by Holwarda \& Bouillaud in the mid-1600s. The availability of electronic computers for astronomical research half a century ago enabled the development of many algorithms to estimate periods quickly and reliably, such as \citet{Lafler1965,Lomb1976,Scargle1982}. 

The aforementioned algorithms work best in the case of periodic variations with constant amplitude and Mira variables present several challenges in this regard. While their periods of pulsation are stable except for a few intriguing cases \citep{Templeton2005}, Mira light curves can exhibit widely varying amplitudes from cycle to cycle \citep[see, for example, the historical light curve of Mira compiled by][]{Templeton2009}. In the case of C-rich Miras, the stochastic changes in mean magnitude across cycles \citep[e.g.,][]{Marsakova1999} only complicate the problem further. The wide variety of light curves for long-period variables, already recognized by \citet{Campbell1925} and \citet{Ludendorff1928}, may complicate the identification of Miras among other stars. Lastly, from a purely practical standpoint, it is simpler to obtain light curves spanning several cycles for RR Lyraes or Cepheids (with periods ranging from $\sim 0.5$ to $\sim 100$~d) than for Miras (with periods ranging from $\sim 100$ to $\sim 1500$~d).

Despite these challenges, the identification and determination of robust periods for Miras --- especially in the regime of sparsely sampled, low signal-to-noise light curves --- would be very beneficial for the determination of distances to galaxies of any type. Thanks to the unprecedented temporal coverage of the Large Magellanic Cloud (LMC) by microlensing surveys, the availability of large samples of extremely well-observed Miras has led to a thorough characterization of their period-luminosity relations at various wavelengths \citep{Wood1999,Ita2004,Soszynski2007}. The dispersion of the $K$-band period-luminosity relation \citep[$\sigma=0.13$~mag]{Glass2003}, is quite comparable to that of Cepheids at the same wavelength \citep[$\sigma=0.09$~mag]{Macri2015} and makes them competitive distance indicators.

\begin{figure}[htbp]
\label{fig:example.mira.lc}
\includegraphics[width=0.49\textwidth]{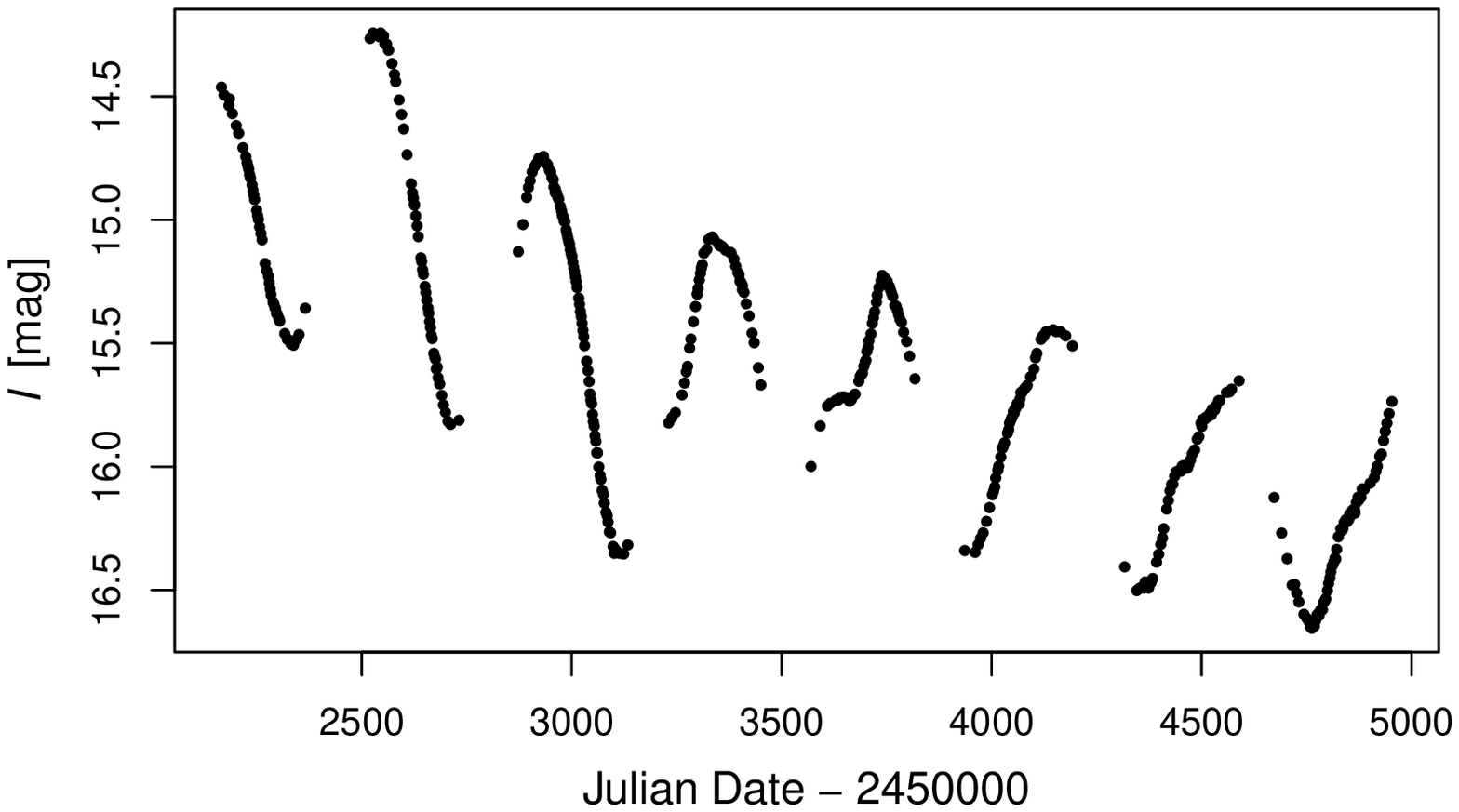}
\includegraphics[width=0.49\textwidth]{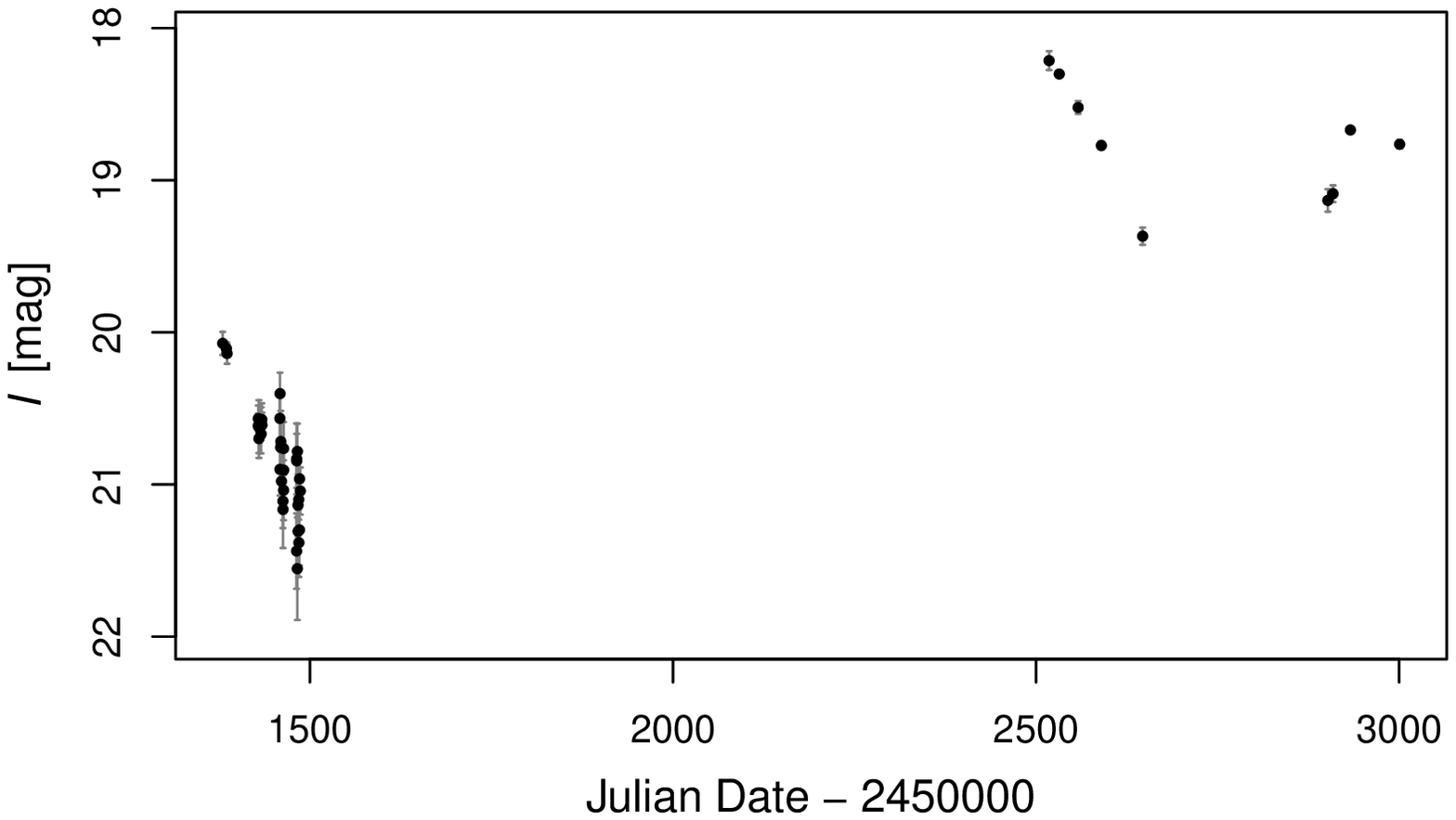}
\caption{Representative Mira light curves observed by OGLE-III in the Large Magellanic Cloud (top) and DIRECT/M33SSS in M33 (bottom).}
\vspace*{-12pt}
\end{figure}

The third phase of the OGLE survey \citep{Udalski2008} imaged most of the LMC with little interruption over 7.5 years and resulted in the discovery of 1663 Miras \citep{Soszynski2009} with a median of 466 photometric measurements per object. The temporal sampling of these light curves and their photometric precision are exceptional relative to typical astronomical surveys and make period estimation relatively easy. In comparison, a similar span of observations of M33 by the DIRECT \citep{Macri2001} and M33SSS projects \citep{Pellerin2011} in the $I$-band consists of a median number of 44 somewhat noisy measurements, heavily concentrated in a few observing seasons. Representative Mira light curves from the OGLE \& DIRECT/M33SSS surveys are shown in Fig.~\ref{fig:example.mira.lc}. There are several reasons for the striking difference in quality between these two data sets. The LMC Miras are among the brightest objects in the OGLE fields, whereas their M33 counterparts are among the faintest in the aforementioned surveys of this galaxy. While the effective exposure times of all these surveys are quite comparable, after taking into account differences in collecting area of their respective telescopes, M33 lies approximately 6.2~mag farther in terms of its $I$-band apparent distance modulus. Furthermore, the main goal of the OGLE project (detection of microlensing events) requires a very dense temporal sampling of the survey fields; this is achieved by using a dedicated telescope and is helped by the fact that the LMC is observable nearly all year long from the site. In contrast, the observations of M33 were carried out using shared facilities (available only a few nights per month) with the primary purpose of studying Cepheids and eclipsing binaries (which do not require exceptionally dense temporal sampling), and the galaxy is only observable all night long for $\sim 1/3$ of the year. Standard period estimation algorithms, which work well for high signal-to-noise, well sampled light curves such as those obtained by OGLE, will fail on more typical data sets represented by the M33 observations. The purpose of this work is to develop and test a methodology for estimating periods for sparsely sampled, noisy, quasi-periodic light curves such as those of Miras observed in M33 by the aforementioned projects.

The rest of the paper is organized as follows. In \S\ref{sec.background} we review several existing period estimation methods. In \S\ref{sec.model} we introduce a new semi-parametric (SP) model for Mira variables which uses a Gaussian process to account for deviations from strict periodicity. We use maximum likelihood to estimate the period and the parameters of the Gaussian process, while other nuisance parameters in the model are integrated out with respect to some prior distributions using earlier studies. Since the likelihood is highly multimodal for the period/frequency parameter, we implement a hybrid method that applies the quasi-Newton algorithm for Gaussian process parameters and a grid search for the period/frequency parameter. In order to assess the effectiveness of the SP model, in \S\ref{sec.construct.test} we carefully construct a simulated data set by fitting smooth functions to the light curves of well-observed OGLE LMC Miras  and resampling them at the cadence, noise level, and completeness limits of the aforementioned M33 observations. Using the simulated data, in \S\ref{sec.evaluation} we compare the performance of existing period estimation methods to our SP model. We find that our proposed model shows an improvement over the generalized Lomb-Scargle (GLS) model under various metrics. In \S\ref{sec.discussion}, we conclude and discuss some future applications. Simulated light curves for reproducing the results in the paper and performance benchmarking are made publicly available as supplementary material. 

\section{Period estimation techniques}\label{sec.background}
Let $y_i$ be the magnitude of a variable star observed at time $t_i$ (in units of days) with uncertainty $\sigma_i$. The data set for this object, obtained as part of a time-series survey with $n$ epochs is $\{(t_i,y_i,\sigma_i)\}_{i=1}^n$. One common approach to estimate the primary frequency of such an object is to assume some parametric model for brightness variation and then use maximum likelihood to estimate parameters. \citet{Zechmeister2009} define the GLS model as
\begin{equation}
\label{eq:gls}
y_i = m + a\sin(2\pi f t_i + \phi) + \sigma_i\epsilon_i,
\end{equation}
where $\epsilon_i \sim \mathcal{N}(0,1)$, $m$ is the mean magnitude, $a$ is the amplitude, $\phi \in [-\pi,\pi]$ is the phase, and $f$ is the frequency \citep[see][for early work in this model]{Reimann1994}. Using the sine angle addition formula and letting $\beta_1 = a\cos(\phi)$ and $\beta_2 = a\sin(\phi)$ one obtains
\begin{equation}
\label{eq:gls2}
y_i = m + \beta_1\sin(2\pi f t_i) + \beta_2\cos(2\pi f t_i) + \sigma_i\epsilon_i.
\end{equation}
The likelihood function of this model is highly multimodal in $f$. However at a fixed $f$ the model is linear in the parameters $(m,\beta_1,\beta_2)$. These two facts motivate the computation strategy of performing a grid search across frequency and minimizing a weighted least squares
\begin{equation}
\begin{split}
& (\widehat{m}(f),\widehat{\beta}_1(f),\widehat{\beta}_2(f) ) \\
& \qquad = \argmin_{m,\beta_1,\beta_2} \sum_{i=1}^n \frac{1}{\sigma_i^2} \left\{y_i - m \right.\\
& \qquad \qquad \left. -\beta_1\sin(2\pi f t_i) - \beta_2\cos(2\pi f t_i)\right\}^2,
\end{split}
\end{equation}
at every frequency $f$ on the grid. Under the normality assumption, the weighted least squares minimization is equivalent to maximizing the likelihood.
Since the model is linear, computation of $\widehat{m}(f),\widehat{\beta}_1(f),\widehat{\beta}_2(f)$ is straightforward. The residual sums of squares at $f$ is 
\begin{equation}
\begin{split}
{\rm RSS}(f) & = \sum_{i=1}^n \frac{1}{\sigma_i^2} \{y_i - \widehat{m}(f) \\
& \quad - \widehat{\beta}_1(f)\sin(2\pi f t_i) - \widehat{\beta}_2(f)\cos(2\pi f t_i)\}^2,
\end{split}
\end{equation}
and the maximum likelihood estimator for $f$ is
\begin{equation}
\widehat{f} = \argmin_{f} {\rm RSS}(f).
\end{equation}
Define ${\rm RSS}_0$ as the (weighted) sum of squared residuals when fitting a model with only an intercept term $m$. The periodogram is defined as
\begin{equation}
\label{eq:glsP}
S_{\rm LS}(f) = \frac{(n-3)({\rm RSS}_0 - {\rm RSS}(f))}{2{\rm RSS}(f)}.
\end{equation}
The periodogram has the property that if the light curve of the star is white noise (i.e., $y_i = m + \epsilon_i$), $S_{\rm LS}(f)$ has an $F_{2,n-3}$ distribution. Thus the periodogram may be used for controlling the ``false alarm probability,'' the potential that a peak in the periodogram is due to noise \citep{Schwarzenberg1996}.

A large number of period estimation algorithms in astronomy are closely related to GLS. The LS method is identical to GLS but first normalizes magnitudes to mean $0$ and does not fit the $m$ term \citep{Lomb1976,Scargle1982}. The ``harmonic analysis of variance'' includes an arbitrary number of harmonics in Equation \eqref{eq:gls} \citep{Quinn1991,Schwarzenberg1996}. \citet{Bretthorst2013} incorporates Bayesian priors on the parameters $\beta_1$ and $\beta_2$. The method is similar to performing a discrete Fourier transform and selecting the frequency which maximizes the \citet{Deeming1975} periodogram. However, \citet{Reimann1994} showed that GLS has better consistency properties than the Deeming periodogram.

\begin{figure}
\centering
\includegraphics[width=0.49\textwidth]{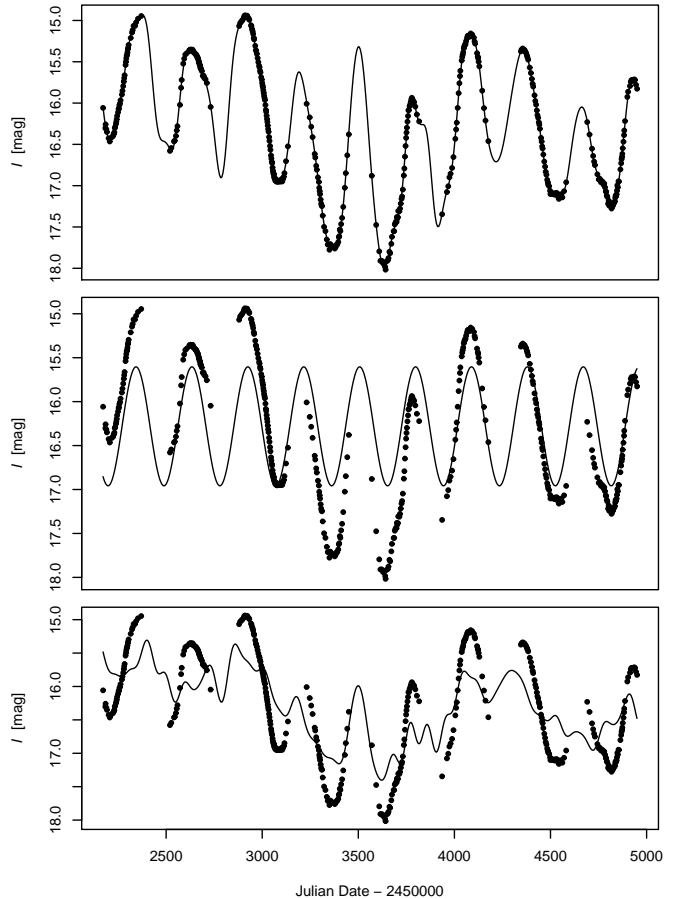}
\caption{Light curve of a Mira in the LMC observed by OGLE (black points), decomposed following Eqn.~\ref{eqn:basicDecomposition}. Top panel: fitted light curve; middle panel: periodic signal, $m+q(t)$; bottom panel: stochastic variations, $m+h(t)$.} \label{fig:onemira}
\vspace*{-12pt}
\end{figure}
 
It is also possible to use non-sinusoidal models but compute and minimize the residual sum of squares as above. For example, \citet{Hall2000} consider the Nadaraya-Watson estimator and \citet{Reimann1994} uses the Supersmoother algorithm. \citet{Wang2012} used Gaussian processes with a periodic kernel and found the period with maximum likelihood or minimum leave-one-out cross-validation error. 

None of the above methods account for the non-periodic variation present in Miras. While these methods are adequate for densely sampled Mira light curves (where the quantity of data overwhelms model inadequacy), their performance deteriorates in the sparsely sampled regime. In Section~\ref{sec.evaluation}, we compare our proposed model with the LS method.

\section{The SP model}\label{sec.model}
Suppose the data $\{(t_i,y_i,\sigma_i)\}_{i=1}^n$ are modeled by $$y_i = g(t_i) + \sigma_i\epsilon_i\, ,$$ where $g(t_i)$ is the light curve signal and the $\epsilon_i\sim N(0,1)$ is independent of other $\epsilon_j$s. The signal of the light curve is further decomposed into three parts, 
\begin{equation}
\begin{split} 
g(t) & = m + q(t) + h(t)\\
     & = m + \beta_1\cos(2\pi ft) + \beta_2\sin(2\pi ft) + h(t)\, ,
\label{eqn:basicDecomposition}
\end{split}
\end{equation}
where $m$ is the long-run average magnitude, $q(t) = \beta_1\cos(2\pi ft) + \beta_2\sin(2\pi ft)$ with frequency $f$ is the exactly periodic signal, and $h(t)$ is the stochastic deviation from a constant mean magnitude, caused by the formation and destruction of dust in the cool atmospheres of Miras. Fig.~\ref{fig:onemira} provides an example of the decomposition for a Mira light curve. The first two terms $m + q(t)$ in Eqn.~\ref{eqn:basicDecomposition} are exactly the same as the GLS model of Eqn.~\ref{eq:gls2}. To simplify notation, we define $\vb_f(t) =(\cos(2\pi ft),\sin(2\pi ft))^T$, so that $q(t) = \vb_f(t)^T\vbeta$. The subscript in $\vb_f(t)$ emphasizes that the basis is parameterized by the frequency $f$. 

An SP statistical model is constructed in Eqn.~\ref{eqn:basicDecomposition} if we assume $h(t)$ is a smooth function that belongs to a reproducing kernel Hilbert space $\mathcal{H}$ with norm $\Vert\cdot\Vert_{\mathcal{H}}$ and a reproducing kernel $K(\cdot,\cdot)$. For this model, if the frequency $f$ is known, we obtain a least squares kernel machine considered in \citet{Liu2007}. Because the frequency is unknown, the response function is nonlinear in $f$. This nonlinearity and the multimodality in $f$ of the residual sum of squares provide additional challenges that require a novel solution.

Besides the additive formulation in Eqn.~\ref{eqn:basicDecomposition}, another possible solution to account for the quasi-periodicity is a multiplicative model such as $g(t)  = m + h(t) q(t)$, where the amplitude of the strictly periodic term $q(t)$ is modified by a smooth function $h(t)$. However, the multiplicative model is more computationally intensive in nature and requires imposing a positive constraint on $h(t)$. As we will show in the following subsections, the $h(t)$ term in the additive model can be easily absorbed into the likelihood function. 
Nevertheless, the multiplicative approach is an interesting alternative approach to model formulation and is open to future study.

\subsection{Equivalent formulations}
Following \S5.2 of \citet{Rasmussen2005}, for fixed $f$, the parameters $m,\beta_1,\beta_2$ and $h(t)$ in Eqn.~\ref{eqn:basicDecomposition} are jointly estimated by minimizing
\begin{equation}
\begin{split}
& \sum_{i=1}^n \frac{1}{\sigma_i^2} [
y_i - m - \beta_1\cos(2\pi ft_i) \\
& \qquad - \beta_2\sin(2\pi ft_i) - h(t_i) ]^2 + \lambda \Vert h(\cdot)\Vert_{\mathcal{H}}^2,
\label{eqn:generalFormulation}
\end{split}
\end{equation}
where $\lambda$ is a regularization parameter. A smoothing/penalized spline model for $h(t)$ is a special case of the general formulation of Eqn.~\ref{eqn:generalFormulation} with a specifically defined kernel; see \S6.3 of \citet{Rasmussen2005}. For fixed $\lambda$, the solution of $h(t)$ is a linear combination of $n$ basis  functions $K(t_i,t)$, $i=1,2,\cdots,n$, by the representer theorem \citep{Kimeldorf1971, OSullivan1986}. It is still left for us to choose the regularization parameter $\lambda$ to balance data fitting and the smoothness of the function $h(t)$.

An equivalent point of view to the above regularization approach is to impose a Gaussian process prior on the function $h(t)$; see \S5.2.3 of \citet{Rasmussen2005}. The benefit of this view is that it provides an automatic method for selecting the regularization parameter $\lambda$. In particular, we can absorb $\lambda$ into the definition of the norm $\Vert \cdot\Vert_{\mathcal{H}}$ and assume the term $h(t)$ in  Eqn.~\ref{eqn:basicDecomposition} follows a Gaussian process, $h(t) \sim \calGP (0,k_{\vtheta}(t,t'))$,  with the squared exponential kernel $k_{\vtheta} (t,t') = \theta_1^2\exp \left(-\frac{(t-t')^2}{2\theta_2^2}\right),$ and parameters $\vtheta = (\theta_1,\theta_2)$. The Gaussian process assumption implies that at any finite number of time points $t_1,t_2,\cdots,t_s$, the vector $(h(t_1),\cdots,h(t_s))$ is multivariate normally distributed, with zero mean and covariance matrix $\mathbf{K} = (k(t_i,t_j))$. This imposes a prior on the function space of $h(t)$. 
We also impose priors on $m$ and $\vbeta$ in Eqn.~\ref{eqn:basicDecomposition}.  In particular, we assume $m \sim \calN(m_0, \sigma_m^2)$ and $ \vbeta \sim \calN(\vzero,\sigma_b^2\I )$.  The prior mean $m_0$ can be interpreted as the average magnitude of Miras in a certain galaxy, and $\sigma_m^2$ is the variance of Miras in that galaxy; the prior variance $\sigma_b^2$ is the variance of the light curve amplitude. These prior parameters can be determined using previous studies. For example, in \S\ref{sec.evaluation}, we use well-sampled light curves of LMC Miras \citep{Soszynski2009} to obtain values of these parameters. It is advisable to check the sensitivity of these prior specifications. 

The benefit of using priors on $m$ and $\vbeta$ is three-fold: first, they introduce regularization by using information from early studies; second, they provide a natural device for separating the estimation of frequency and the light curve signal component using Bayesian integration when the parameter of interest is the frequency; lastly, the regularization parameter $\vtheta$ of the non-parametric function is allowed to be chosen by the maximum likelihood, without resorting to the computationally expensive cross-validation method.

In summary, we have built the following hierarchical model for a Mira light curve:

\begin{equation}
\label{eqn:hier}
\begin{split}
& y_i | m,\vbeta, g(t_i)  \sim \calN(g(t_i), \sigma_i^2), \\
& g(t)  = m + \vb_f(t)^T\vbeta + h(t), \\
& m   \sim \calN(m_0, \sigma_m^2), \vbeta  \sim \calN(\vzero,\sigma_b^2\I ),\\
& h(t)| \vtheta \sim  \calGP(0,k_{\vtheta} (t,t')),
\end{split}
\end{equation}
where $\vtheta$ and $f$ are fixed parameters. In this model, the frequency parameter $f$ is of key interest to our study. We do not perform a fully Bayesian inference by imposing a prior distribution on $f$ because the likelihood function of $f$ is highly irregular, with numerous local maxima, and Monte Carlo computation of the posterior is expensive and intractable for large astronomical surveys.

Previously, \citet{Baluev2013} applied a Gaussian process model to study the impact of red noise in radial velocity planet searches. While his maximum likelihood method is a classical frequentist approach in statistics, our approach can be considered as a hybrid of Bayesian and frequentist approaches. We treat the parameter of interest $f$, and the parameters for the kernel $\vtheta$ of the Gaussian process as fixed, and impose a prior distribution on other parameters. This is similar to the type-II maximum likelihood estimation of parameters of a Gaussian process or regularization parameters in function estimation; see \S5.2 of \citet{Rasmussen2005}. From the Bayesian point of view, $\vtheta$ and $f$ are treated as hyper-parameters that in turn are estimated by the empirical Bayes method. Because the Gaussian process plays a critical role in modeling departure of light curves from periodicity, we may also refer to our model more precisely as the nonlinear SP Gaussian process model.

\subsection{Estimation of the frequency and the periodogram}
Let $\vy = (y_1,y_2,\cdots, y_n)$ be the observation vector of the magnitudes of a light curve. By integrating out $m,\vbeta$ and $\vh$ from the joint distribution given by Eqn.~\ref{eqn:hier}, we get the marginal distribution of $\vy$, $p(\vy|\vtheta, f)$, which is a multivariate normal with mean $\vmu= m_0\vone$ and covariance matrix 
$$\K_y = \left( \sigma_m^2 + \sigma_b^2 \vb_f(t_i)^T \vb_f(t_j) +
 k_{\vtheta}(t_i, t_j) +\sigma^2_i\delta_{ij} \right)_{n\times n}, $$
where $\delta_{ij} = 1$ if $i=j$ and $\delta_{ij} = 0$ if $i\neq j$. 
 Therefore, the log likelihood of $\vtheta$ and $f$ is
\begin{equation}
\begin{split}
Q(\vtheta,f) = &\log(p(\vy| \vtheta, f))\\
             = &-\frac{1}{2} (\vy-m_0\vone)^T\K_y^{-1}(\vy-m_0\vone) \\
               &\qquad -\frac{1}{2}\log\det \K_y -\frac{n}{2}\log(2\pi) \label{eqn:mainObj}.
\end{split}
\end{equation}

The maximum likelihood estimator of $\vtheta$ and $f$ is obtained by maximizing $Q(\vtheta,f)$. Since the likelihood function is differentiable with respect to $\vtheta$ but  highly multimodal in the parameter $f$, standard optimization methods cannot be directly used to jointly maximize over $\vtheta$ and $f$.

We adopt a profile likelihood method as follows. For each frequency $f$ over a dense grid, we compute the maximum likelihood estimator $\widehat{\vtheta}_f = \argmax_{\vtheta} Q(\vtheta,f)$. This can be done using the quasi-Newton method. Then we perform a grid search to find the maximum profile likelihood estimator of $f$, i.e., 
\begin{equation}
\label{eqn:fhat}
\hat{f} = \argmax_f Q(\widehat{\vtheta}_f, f)\, , 
\end{equation}
the estimated period is $\hat{P}=1/\hat{f}$. The details of the algorithm are given in \S\ref{sec.quasi}. The profile log-likelihood as a function of the frequency $f$ is adopted as the \textit{periodogram} for our model,
\begin{equation}
S_{SP}(f) = Q(\widehat{\vtheta}_f, f)\, . \label{eqn:periodogram}
\end{equation}
It contains the spectral information of the signal. The frequency of the dominant harmonic component is expected to be the location of the peak of this profile likelihood. 

\subsection{Computation of the periodogram}\label{sec.quasi}
\begin{figure}[t]
\centering
\includegraphics[width = 0.43\textwidth]{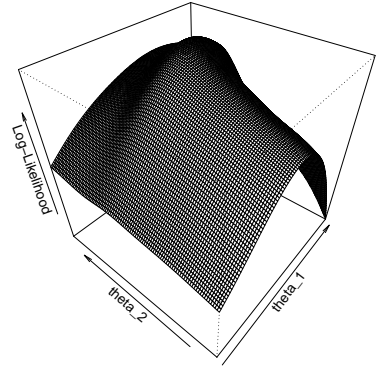}
\caption{The three dimensional surface plot of $Q(\vtheta, f)$ in Eqn.~\ref{eqn:mainObj}, for the simulated light curve in Fig.~\ref{fig:lc.and.spec}. Notice $Q(\vtheta, f)$ is plotted as a function of $\vtheta$ and $f$ is fixed at its true frequency.}\label{fig:lc.surface}
\end{figure}

Now we present the details of computing the profile likelihood. Because $Q(\vtheta,f)$ is highly multimodal in the frequency parameter $f$, we follow the commonly used strategy of optimization through grid search. On the other hand, since $Q(\vtheta,f)$ is differentiable in parameter $\vtheta$, the quasi-Newton method can be employed to optimize over $\vtheta$ for fixed $f$, and obtain the profile likelihood (Eqn.~\ref{eqn:periodogram}). The gradient of the log likelihood (Eqn.~\ref{eqn:mainObj}) with respect to $\theta_j (j=1,2)$ is
\[\frac{\partial}{\partial \theta_j} Q(\vtheta,f)\!=\! \frac{1}{2}\mathrm{tr}\left((\valpha\valpha^T\!-\!\K_y^{-1})\frac{\partial \K_y}{\partial \theta_j}\right)\] where $\valpha\!=\!\K^{-1}_y(\vy-m_0\vone)$. In general, the objective function for the Gaussian process model is not convex in its kernel parameters $\vtheta$ and global optimization cannot be guaranteed. Fig.~\ref{fig:lc.surface} shows a surface plot of $Q(\vtheta,f)$ as a function of $\vtheta$ for one simulated light curve, with $f$ fixed at the true frequency. The surface exhibits unimodality in this case, although it is not convex.

The computation involved in calculating the profile likelihood through the quasi-Newton method can be intensive. Since the objective function (Eqn.~\ref{eqn:mainObj}) is non-convex in $\vtheta$, generally multiple starting points should be attempted to find the global optimizer when applying the quasi-Newton method. In addition, evaluating the objective function and the gradient function requires inversion of the covariance matrix whose computation cost is of the order $O(n^3)$. During each quasi-Newton iteration, these evaluations could be repeated several times because multiple step size might be attempted. To make the computation more challenging, all of the above needs to be repeated at hundreds or even thousands of densely gridded $f$s per light curve. Furthermore, the method may need to be applied to hundreds of thousands or millions of light curves from large astronomical surveys. 

In order to speed up computation over the dense grid of frequency values, we use the result of applying the quasi-Newton method at one frequency value as a warm start for the subsequent frequency value. Specifically, the optimizer $\widehat{\vtheta}_f$ and its approximate inverse Hessian matrix are provided as quantities to start the quasi-Newton iterations for the next frequency value on the dense grid. When the initial point is near the local minimizer and the inverse Hessian matrix is a good approximation to the true Hessian matrix, the quasi-Newton algorithm will converge at superlinear rate; the step size of $\alpha=1$ will be accepted by the Wolfe descent condition, avoiding evaluation of the objective function multiple times to determine the appropriate step size during each iteration \citep[see Ch.~6 of][for a more rigorous mathematical discussion]{Nocedal2006}. We find that a warm start can speed up the computation significantly but sometimes we need to restart with random initial values to ensure convergence to the global optimum. The pseudocode provided in the Appendix describes our algorithm.

\subsection{Estimation of the signal and its components}
After the parameters $f$ and $\vtheta$ are fixed at their maximum likelihood estimates $\widehat{f}$ and $\widehat{\vtheta}_{\hat{f}}$, we can perform the inference of the light curve signal $g(t)$ and its components in the standard Bayesian framework. Interested readers may consult Ch.~2 of \citet{Rasmussen2005} for a detailed discussion of this topic.

 Firstly, we could obtain the posterior  distribution of $\vgamma = (m,\vbeta^T)$, the parameters for the long run average magnitude and the exactly periodic term. The prior of $\vgamma$ is $\calN(\vgamma_0,\bSigma_\gamma)$ with $\vgamma_0=(m_0,0,0)^T$ and $\bSigma_\gamma = \mathrm{diag}(\sigma_m^2,\sigma_b^2,\sigma_b^2)$. Its posterior distribution is $\vgamma| \vy \sim  \calN(\bar{\vgamma}, \bar{\bSigma}_\gamma)$ with
\begin{equation}
\label{equ:gammapost}
\begin{split}
\bar{\vgamma}  = & \left(\H^T\K_c^{-1}\H+\bSigma_\gamma^{-1}\right)^{-1} \\
                 & \left(\bSigma_\gamma^{-1}\vgamma_0+\H^T\K_c^{-1}\vy\right)\, , \\
\bar{\bSigma}_\gamma  = &  \left(\H^T\K_c^{-1}\H + \bSigma_\gamma^{-1}\right)^{-1}, 
\end{split}
\end{equation}
\noindent{where} $$\vh(t)\!=\!(1,\vb_{\hat{f}}(t)^T)^T, \H\!=\!(\vh(t_1),\vh(t_2),\cdots,\vh(t_n))^T,$$
and$\,\K_c\!=\!\big(k_{\widehat{\vtheta}_{\hat{f}}}(t_i,\!t_j)\!+\!\sigma_i^2\delta_{ij}\!\big)\!_{{\tiny\textit{n}}\times\!{\tiny\textit{n}}}\,$with$\,\widehat{f}\,$and$\,\widehat{\vtheta}_{\hat{f}}\,$plugged in.

Consider the prediction of light curve magnitude at a specific time point $t^*$. Define the vector $\vk^* = (k_{\widehat{\vtheta}}(t^*,t_1),\ \cdots,\ k_{\widehat{\vtheta}}(t^*,t_n))^T$. Conditional on $(\vy, \vgamma)$, the distribution of  $g(t^*)| \vy, \vgamma$ is a multivariate normal with mean $ \vh(t^*)^T \gamma + \vk_{\vtheta}(t^*,\vt) \K_c^{-1} (\vy - \H\gamma)$ and variance $k_{\widehat{\vtheta}}(t^*,t^*) - (\vk^*)^T \K_c^{-1}  \vk^*$. With the posterior distribution of $\vgamma$ given in  Eqn.~\ref{equ:gammapost}, we are able the remove $\vgamma$ from the above conditional distribution of $g(t^*)$. Finally, we get the posterior distribution of the signal at $t^*$ as $g(t^*)| \vy \sim \calN(\bar{g}^*, \bar{\sigma}_{g^*}^2)$ with
\begin{equation}
\begin{split}
\bar{g}^* = & \vh(t^*)^T \bar{\gamma} + \vk(t^*,\vt) \K_c^{-1}
            (\vy - \H\bar{\gamma})\, ,\\
\bar{\sigma}^2_{g^*} = & k_{\widehat{\vtheta}}(t^*,t^*) - (\vk^*)^T \K_c^{-1}
                         \vk^*+
\vrt^T\bar{\bSigma}_\gamma\vrt\, , 
\end{split}
\label{equ:gpredict}
\end{equation}
where $\vrt=\vh(t^*)-\H^T\K_c^{-1}\vk^*$.

\section{Simulation of M33 light curves}\label{sec.construct.test}
It is not possible to evaluate the period estimation accuracy of our method directly on the M33 data because the ``ground truth'' is unknown. Instead, we construct a test data set by smoothing the well-sampled OGLE light curves to infer continuous functions, then resample these functions to match the observational patterns of the M33 data, and at last add noise to the light curves. This data set can serve as a testbed for future studies of comparing different period estimation methods. We will now describe the M33 observations and the construction of the test data set. As the whole simulation procedure is a complicated process, we will discuss its components in detail from \S4.1 to \S4.4. The whole simulation procedure will be summarized in \S4.5.

\subsection{Characteristics of the M33 observations}
\begin{figure}
\centering
\includegraphics[width = 0.49\textwidth]{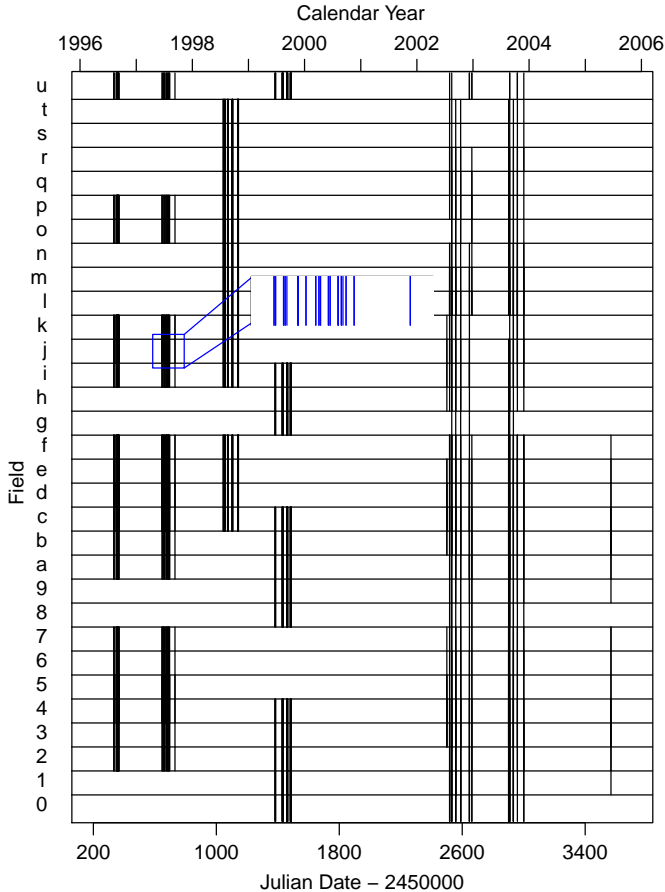}
\caption{Observation patterns for 31 fields in M33, labeled as 0, 1, $\dots$, 9, a, b, $\dots$, u. The horizontal axis shows the Julian date (bottom) and the calendar year (top).}\label{fig:obs.gaps}. 
\vspace*{-18pt}
\end{figure}

\begin{figure}
\centering
\includegraphics[width=0.49\textwidth]{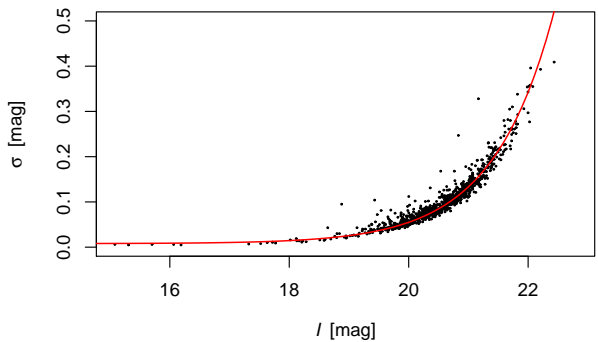}
\caption{$m-\sigma$~relation for a given night and field within M33. The solid red line is the best-fit relation using the empirical function $\sigma = a^{(m-b)} + c$, with $a=2.666,\,b=23.117,\,c=0.008$.}\label{fig:simu.sigmag}
\end{figure}

Most of the disk of M33 was observed by the DIRECT \citep{Macri2001} and M33SSS \citep{Pellerin2011} projects in the $BVI$ bands, with a combined baseline of $7-9$ years and a sampling pattern that depends on the exact location within the disk (see Fig.~\ref{fig:obs.gaps}). The large area of coverage and long baseline of these observations make them suitable for Mira searches. We use the $I$-band observations to carry out the simulations, as this is the wavelength range where Miras are brightest (out of the three bands used by these projects). Detailed descriptions of the M33 observations can be found in the above referenced papers. We use the data products from a new reduction that will be presented in a companion paper (W.~Yuan et al.~2016, in prep.). $I$-band light curves are available for $\sim 2.5\times 10^5$~stars, with a median of 44 measurements and a maximum of 170.
We model the relation between a magnitude measurement $m$ and its uncertainty $\sigma$ as
\begin{equation}
\sigma = a(t_i',F)^{[m-b(t_i',F)]} + c(t_i',F)\, , \label{equ.sigma.mag}
\end{equation}
for each observation field $F$ and each observation night $t_i'$, where $a(t_i',F)$, $b(t_i',F)$ and  $c(t_i',F)$ are field- and night-specific constants. There are 31 different fields in total, $F=0,1,\cdots, 9, a,b,\cdots, u$. The parameters are determined via least-squares fitting using all the measurements for the specific field $F$ and night $t_i'$. Fig.~\ref{fig:simu.sigmag} shows the $m-\sigma$~relation for a typical field.

In order to test the SP periodogram we need sparsely sampled, moderately noisy Mira light curves with known periods. Thus, we characterize the sampling patterns and noise levels of the M33 observations and simulated Mira light curves of known periods using the OGLE observations of these objects in the LMC.

\subsection{Matching  the M33 observation pattern}
The first step in simulating a Mira light curve is to randomly select a sampling pattern based on the  light curve of an actual star in some field $F$, $\{t_i'\}_{i=1}^n$ with $n\in [10,170]$. A random time shift $s$ is added, $t_i = t_i' + s$ for $i=1,2,\cdots, n$. The random shift $s$ follows a uniform distribution over the interval $[0, P_0]$, where $P_0$ is the true period of the LMC Mira selected during the artificial light curve generation process. This helps to simulate a large number of unique light curves sampled at random phases using the limited number of template light curves. 

\subsection{The Mira template  light curves}
\begin{figure}
\centering
\includegraphics[width=0.49\textwidth]{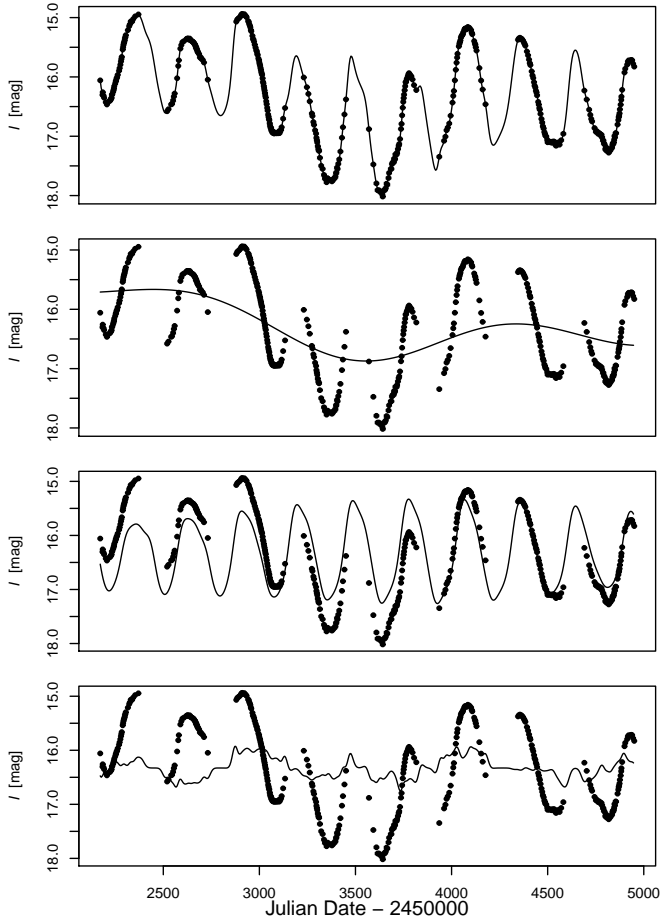}
\caption{Light curve of a Mira in the LMC observed by OGLE (black points), decomposed following Eqn~\ref{eqn:fullgpmodel}. Top panel: the fitted light curve; second panel: long-term signal, $m+l(t)$; third panel: periodic term, $m + q(t)$; bottom panel: stochastic variations, $m+h(t)$.} \label{fig:complexDecomposition}
\vspace*{-12pt}
\end{figure}

The template Mira light curves are obtained by using our SP model to fit the Mira light curves in the LMC, collected by the OGLE project \citep{Soszynski2009}. A total number of 1663 Miras have been observed in $I$ with very high accuracy, excellent phase coverage, and a long baseline (the median and mean number of observations are 466 and 602, respectively, with a baseline of $\sim 7.5$~years for most fields).  Because the LMC light curves are densely sampled with high quality, we can adopt a more complicated model to provide a higher fidelity fit. Following \S5.4.3 of \citet{Rasmussen2005}, instead of Eqn.~\ref{eqn:basicDecomposition}, the signal light curve $g(t)$ is decomposed into 
\begin{equation} \label{eqn:fullgpmodel}
g(t)  = m + l(t) + q(t) + h(t),
\end{equation}
where $m$ is the long run average magnitude, $l(t)$ is the long-term (low-frequency) trend across different cycles, $q(t)$ is the periodic term, and $h(t)$ is small-scale (high-frequency) variability within each cycle. The latter three terms are modeled by the Gaussian process with different kernels. In particular, we use the squared exponential kernel $k_l(t_1,t_2) = \theta_1^2 \exp(-\frac{1}{2}\frac{(t_1-t_2)^2}{\theta_2^2})$ for $l(t)$, another squared exponential kernel $k_h(t_1,t_2) = \theta_6^2 \exp(-\frac{1}{2}\frac{(t_1-t_2)^2}{\theta_7^2})$ for $h(t)$, and lastly a periodic kernel
\begin{equation*}
\begin{split}
k_q(t_1,t_2) = \theta_3^2  \exp \bigg ( -\frac{1}{2}&\frac{(t_1-t_2)^2}{\theta_4^2}  \\
 &- \frac{2\sin^2(2\pi f(t_1-t_2))}{\theta_5^2} \bigg )
\end{split}
\end{equation*}
for $q(t)$. Note the periodic kernel allows the light curve amplitude to change across cycles. The maximum likelihood method is applied to fit each LMC light curve, fixing $f$ to the OGLE value and solving for the unknown parameters $(\theta_1,\theta_2, \cdots, \theta_7)$. Fig.~\ref{fig:complexDecomposition} is an illustration of the model fitting result using Eqn~\ref{eqn:fullgpmodel} based on the same light curve as in Fig.~\ref{fig:onemira}. Notice that the more complex model in Fig.~\ref{fig:complexDecomposition} is only suitable for a densely sampled light curve.

Once the sampling pattern is chosen, one of the template light curves will be selected according to the luminosity function described in the next subsection.  With the selected template, the magnitude of the simulated light curve signal at $t_i'$ with shift $s$ is $g(t_i'+s)$, which is computed with Eqn.~\ref{eqn:fullgpmodel} in a similar way as Eqn.~\ref{equ:gpredict}. 

\subsection{Matching the luminosity function\\to the M33 observations}
While the OGLE observations of LMC Miras are deep enough to detect these objects over their entire range of luminosities, the M33 observations become progressively more incomplete for fainter and redder objects. We derived an empirical completeness function for the M33 observations as follows. We fitted the observed luminosity function $\mathcal{F}_0(I)$ using an exponential for $I \in [18.5,20]$~mag and extrapolated to fainter magnitudes, obtaining $\mathcal{F}_1(I)$. The empirical completeness function is then $\mathcal{C}(I) = \mathcal{F}_1(I) / \mathcal{F}_0(I)$. 

We randomly picked $\{t_i'\}_{i=1}^n$ from the M33 light curves. For each $\{t_i'\}_{i=1}^n$, we selected a (LMC-based) template using $\mathcal{C}(I+6.2)$ as the probability distribution. The value of $+6.2$~mag accounts for the approximate difference in distance modulus between the LMC and M33. In this way the resulting luminosity function of the simulated light curves is statistically the same as that of the real M33 observations.

\subsection{The simulation procedure}
With all the components discussed above, we are able to present the whole simulation procedure here. In order to generate one simulated Mira light curve matching the sampling characteristics of the M33 observations, the first step is to randomly select a sampling pattern $\{t_i'\}_{i=1}^n$, and then add a random shift $s$, $t_i = t_i' + s$, $i=1,2,\cdots, n$. The second step is to randomly select a template light curve according to the luminosity function, then compute the light curve signal $g(t_i'+s)$ for the selected sampling pattern $\{t_i'\}_{i=1}^n$. The third step is to use the best-fit relations (Eqn.~\ref{equ.sigma.mag}) to add photometric noise via $$ y_i = g(t_i' + s)+6.2+ \sigma_i\epsilon_i\, ,$$ where $+6.2$~mag is the approximate relative distance modulus, $\epsilon_i$ is drawn from $\mathcal{N}(0,1)$, and $\sigma_i$ is computed from $$\sigma_i = a(t_i',F)^{[g(t_i)+6.2-b(t_i',F)]} + c(t_i',F). $$ for the selected observation pattern $t_i'$ and field $F$. Following this procedure, we generate one simulated light curve $\{t_i',y_i,\sigma_i\}_{i=1}^n$. The procedure is repeated until $10^5$ suitable light curves are generated, excluding any with $<10$~data points or sampling on $<7$ nights.

\section{Performance evaluation}\label{sec.evaluation}
Having generated the test data set, we evaluate the performance of the SP model and compare it with the GLS model. We choose prior parameters for the SP model of $m_0 = 15.62 + 6.2$, $\sigma_m = 10$ and $\sigma_b =1$. The adopted value of $m_0$ is the average $I$ magnitude of Miras in the LMC and once again $+6.2$ is the approximate relative distance modulus between M33 and the LMC. The values of $\sigma_m$ and $\sigma_b$ are larger than those derived from the LMC samples in order to make those priors non-informative. Although fitting the SP model is computationally slower than the LS model, we find that our model gives an overall improvement in various metrics. For both methods, the periodograms are computed on a dense frequency grid from $1/2000$ to $1/100$ with a spacing of the order of $10^{-5}$. For the GLS method, we chose a spacing of (0.05/time span) or $\sim2.5\times 10^{-5}$, which results in optimal performance for this simulation. For our SP method, we chose a slightly smaller value of $10^{-5}$ to facilitate the warm start mechanism in our algorithm (see Appendix) given that small changes in frequency result in tiny changes of the objective function.
 
\subsection{The aliasing effect}

\begin{figure}[t]
\centering
\includegraphics[width = 0.43\textwidth]{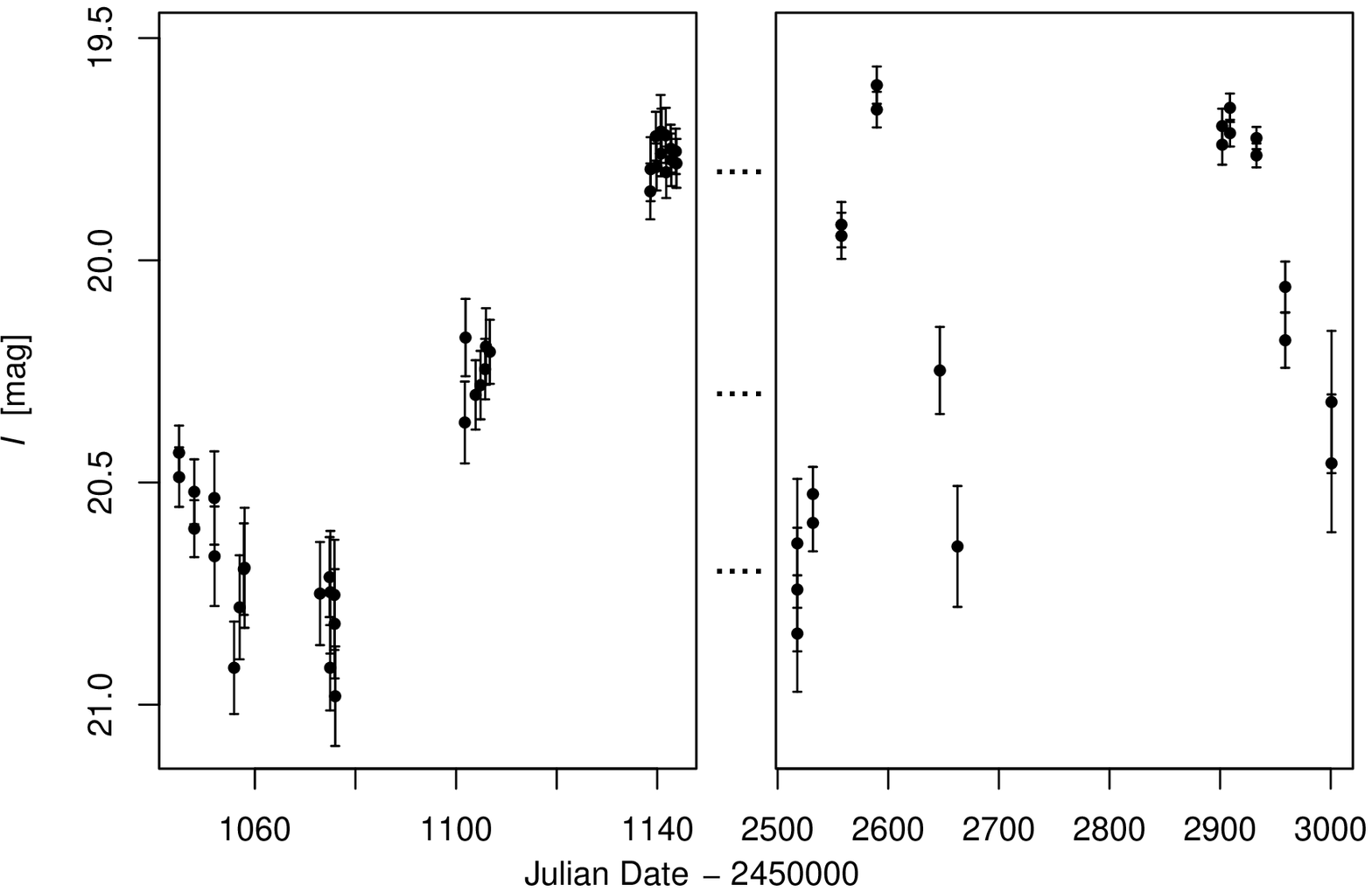}
\includegraphics[width = 0.43\textwidth]{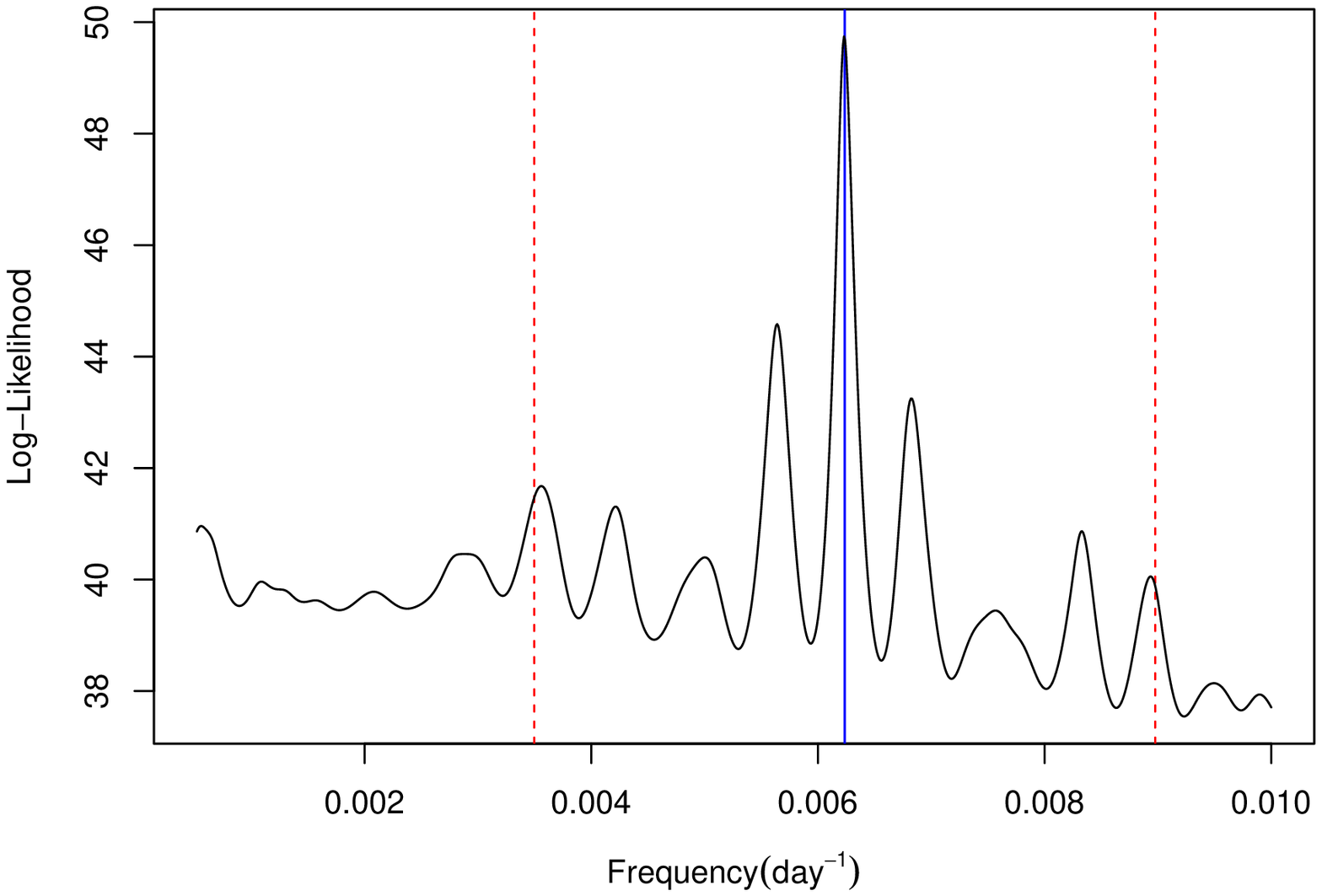}
\caption{Simulated light curve (top) and periodogram $S_{SP}(f)$ based on our model (bottom). Error bars are derived from the M33 observations. In the top panel, a large gap in temporal coverage was removed to make the plot compact (also note the different time spans). The lines in the bottom panel correspond to the true (solid blue) and the one-year aliasing (dashed red) frequencies.}\label{fig:lc.and.spec}
\vspace*{-6pt}
\end{figure}

We fit the entire simulated data set using the SP model. Fig.~\ref{fig:lc.and.spec} gives an example of a simulated light curve and its SP periodogram (Eqn.~\ref{eqn:periodogram}). In this example, the true frequency (labeled by the blue dotted line) is successfully recovered. 

Aliasing frequencies at $f\pm 1/365$~d affect most periodograms when dealing with sparsely observed astronomical data. The red dashed line in Fig.~\ref{fig:lc.and.spec} indicates the aliasing frequency at $f+1/365$ where a strong peak exists. This is not a rare case, and for some light curves the one-year beat aliasing frequencies have higher log likelihoods than the true frequencies. Fig.~\ref{fig:fvsf} compares the recovered and true frequencies for all simulated light curves. Two secondary strips parallel to the main one and offset by $\pm0.00274$ represent $\hat{f} = f \pm 1/365$, respectively. Other aliasing frequencies, such as $2f$, $3f$, $0.5f$, etc., are also noticeable. Lastly, due to the sampling pattern of some light curves, the side lobes of the main peak can be higher than the central value. These manifest as close parallel strips to the aforementioned features.

\subsection{Accuracy assessment}

\begin{figure}[htb]
\includegraphics[width = 0.48\textwidth]{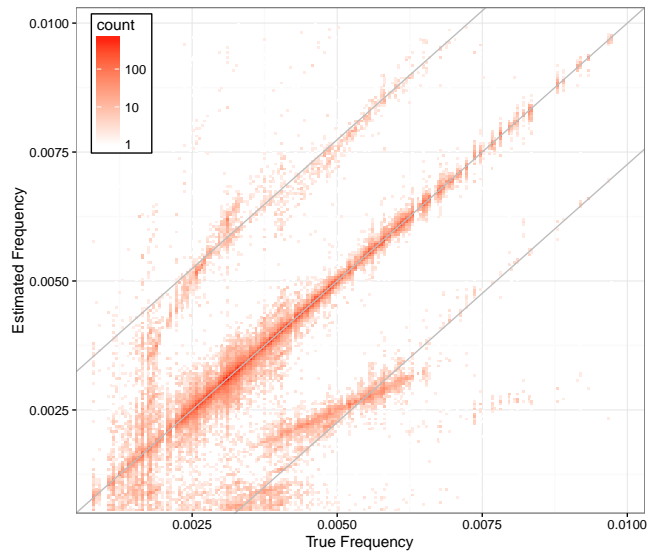}
\caption{Model-recovered vs.~true frequencies of the test data set. Features other than the one-to-one line are due to aliasing from one-year beat or harmonic frequencies. The three gray diagonal lines correspond to $\hat{f} = f + 1/365, f $ and $f - 1/365$ from top to bottom, respectively.}\label{fig:fvsf}
\vspace*{-12pt}
\end{figure}

\begin{figure*}[hbtp]
\centering
\includegraphics[width = 0.32\textwidth]{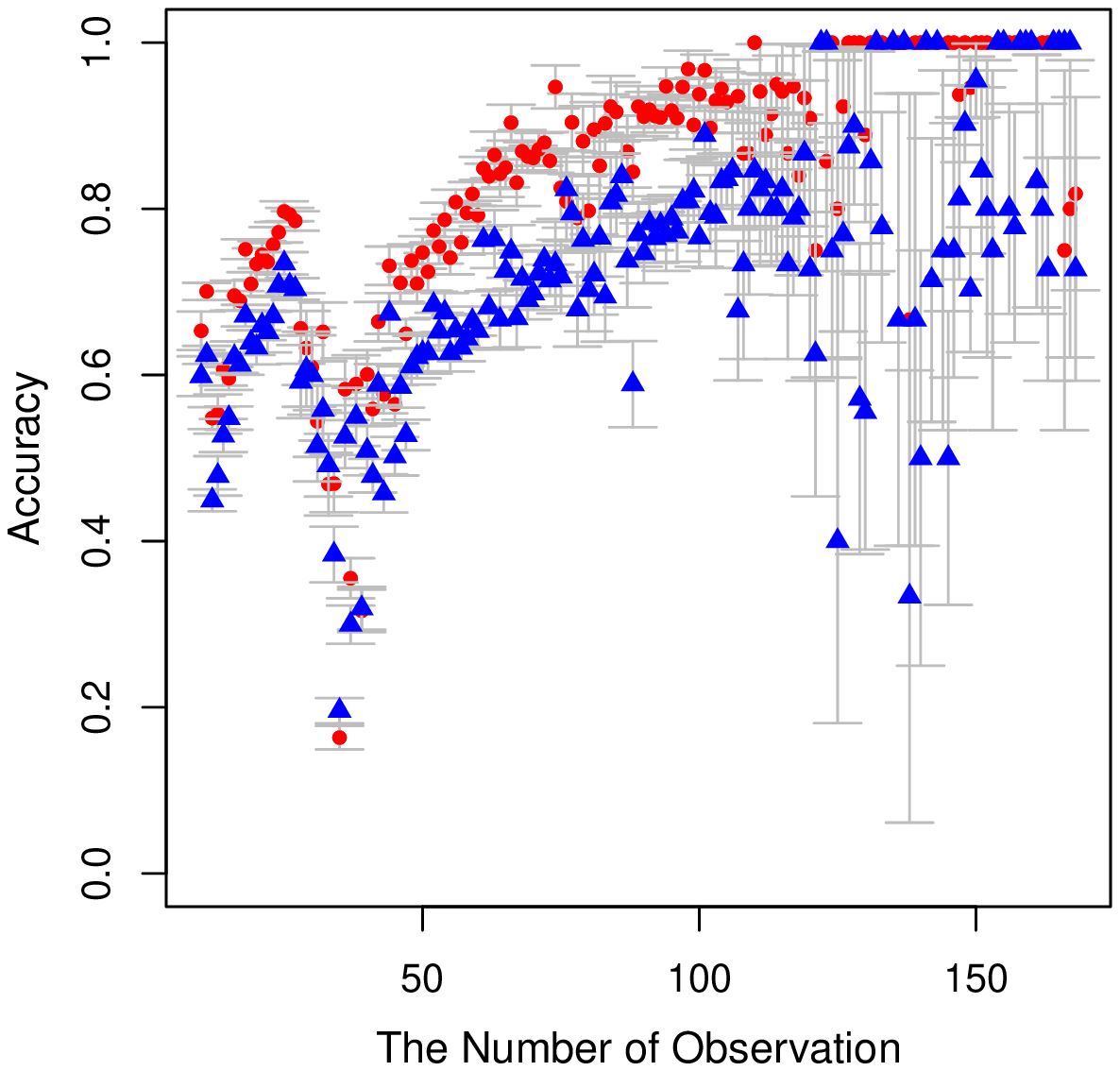}
\includegraphics[width = 0.32\textwidth]{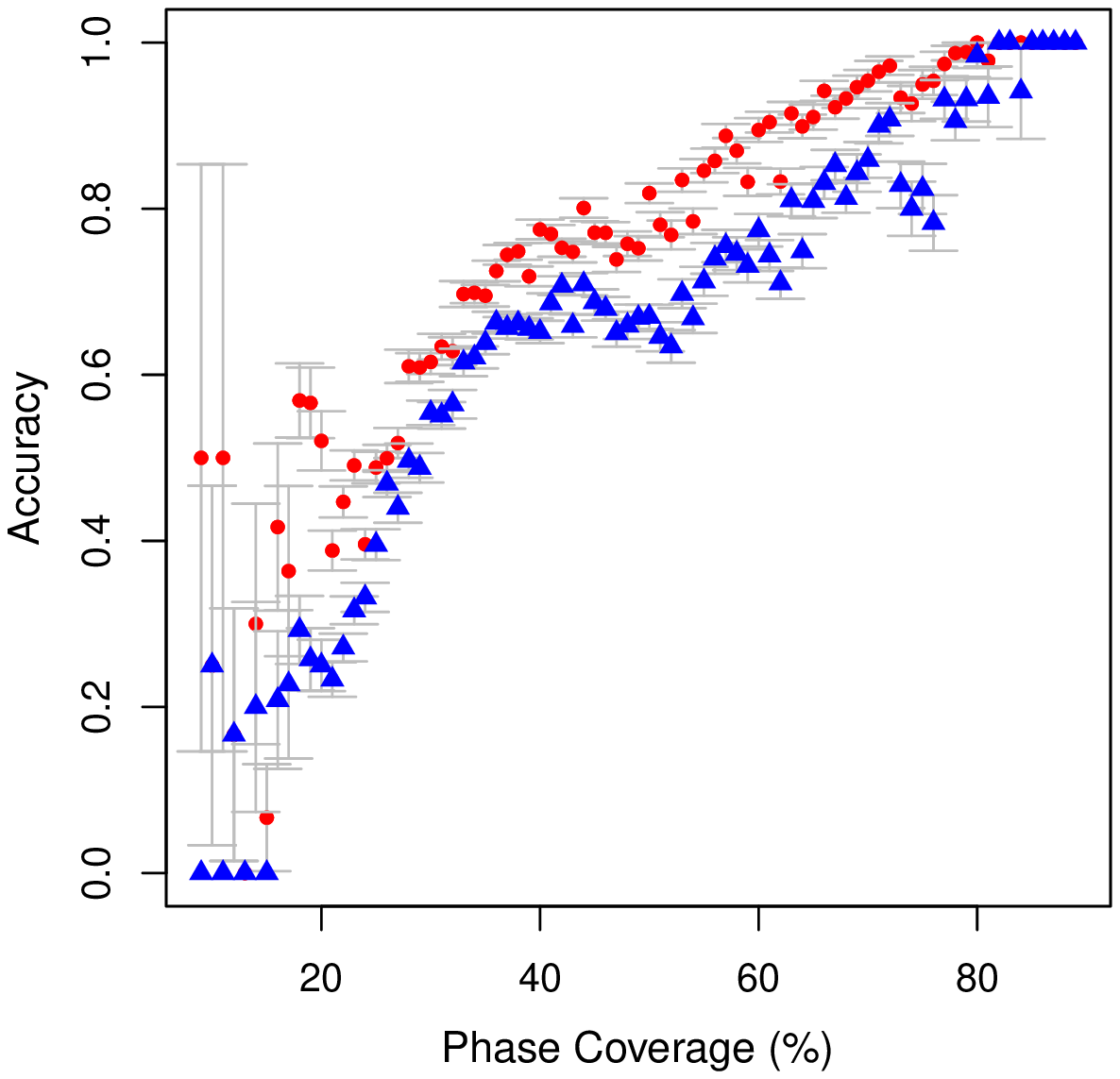}
\includegraphics[width = 0.32\textwidth]{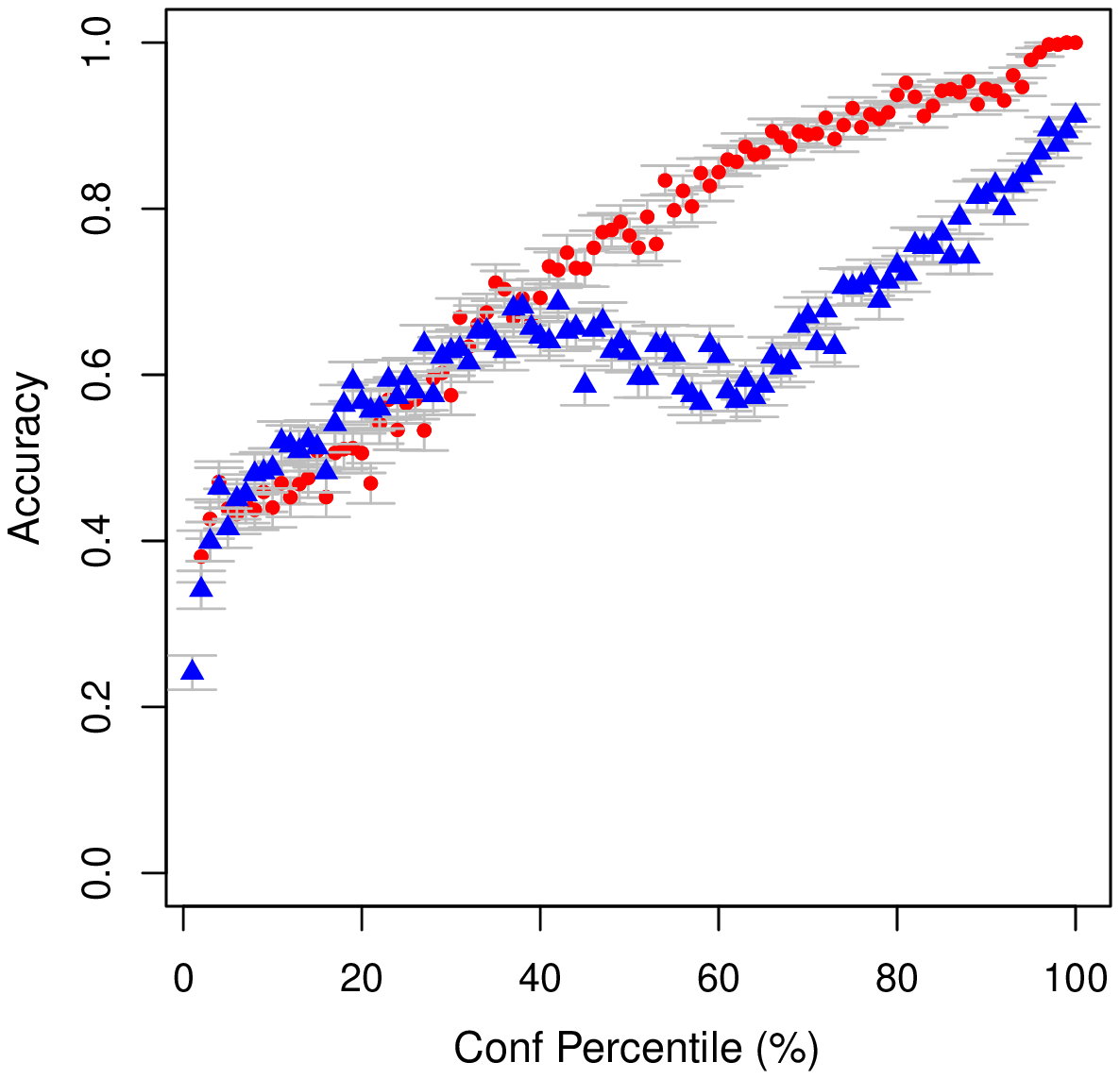}
\includegraphics[width = 0.32\textwidth]{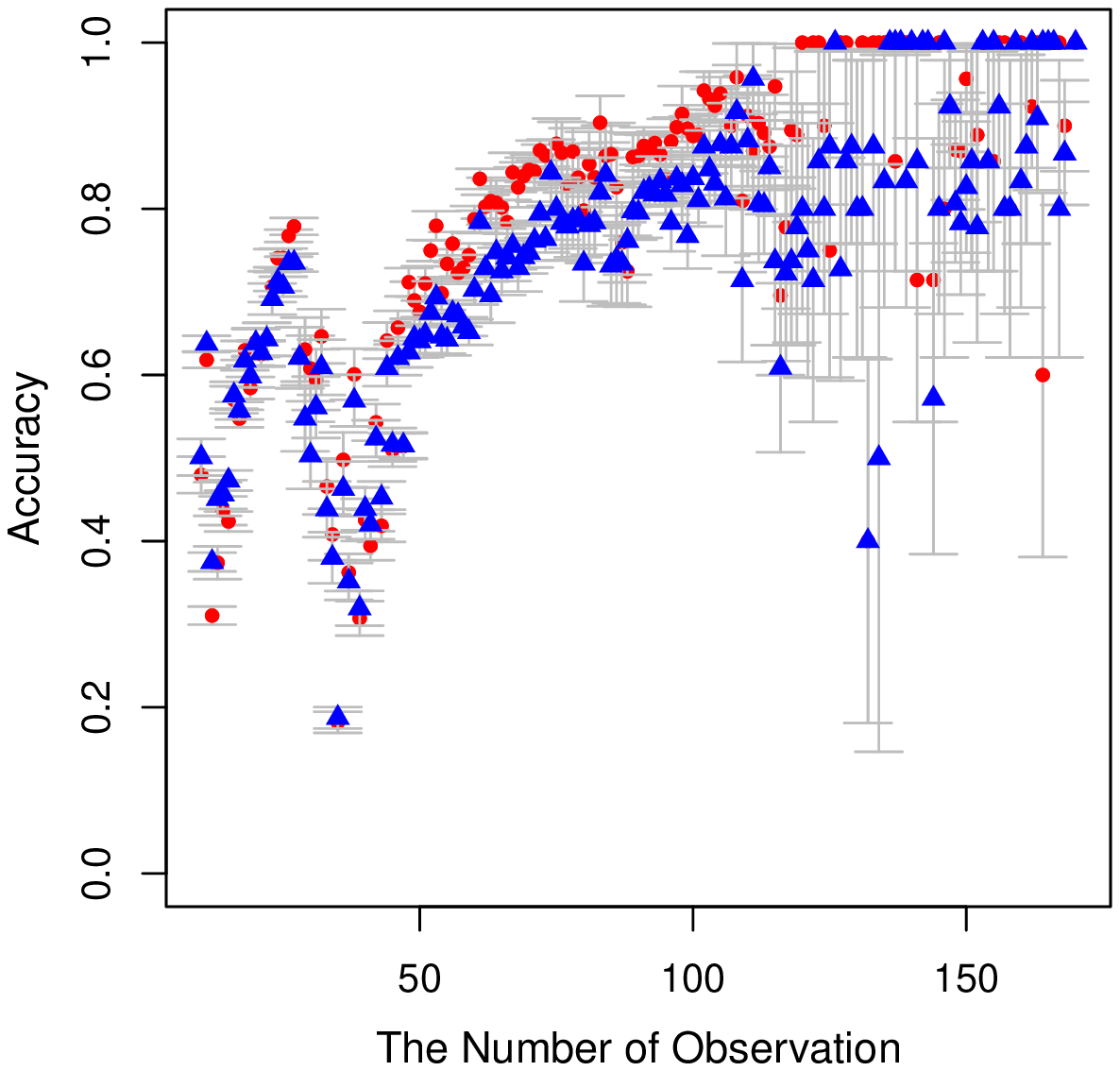}
\includegraphics[width = 0.32\textwidth]{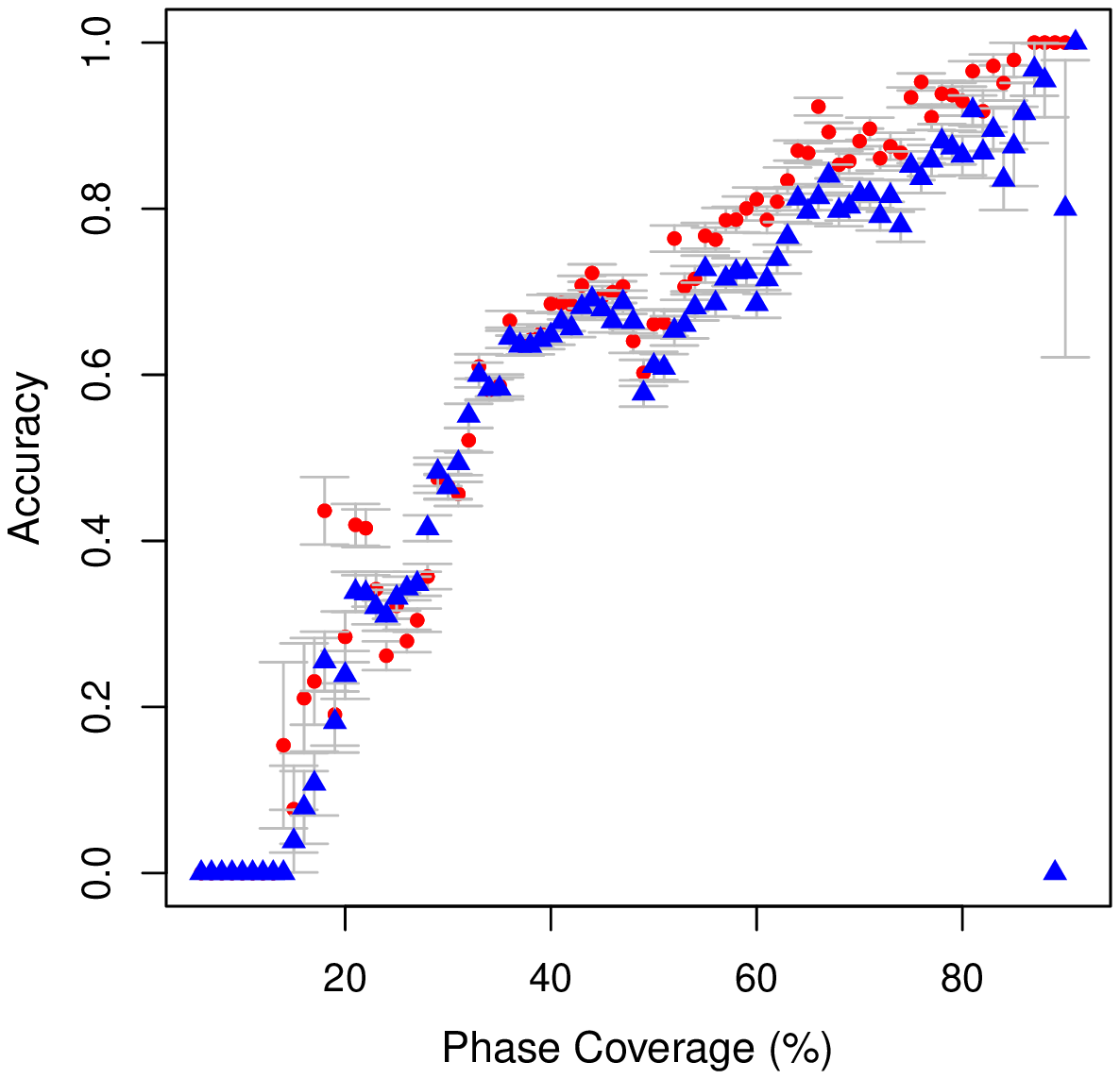}
\includegraphics[width = 0.32\textwidth]{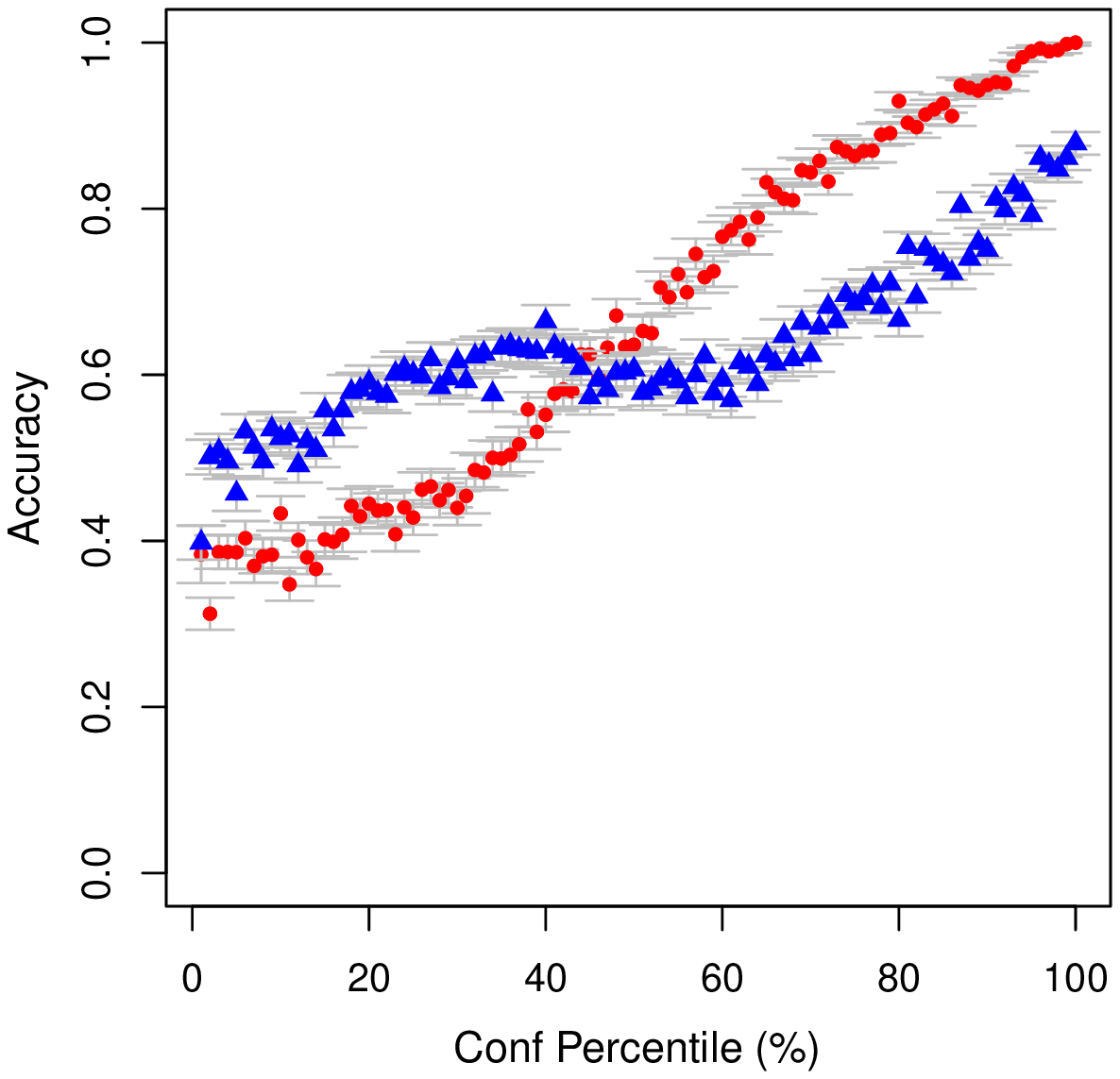}
\caption{Accuracy comparison using different metrics. Light curves are grouped by number of observations (first column), phase coverage (second column), and {\tt conf} values (third column). The estimation accuracies in each group is computed and plotted as above. The red circles represent the SP model, while blue triangles denote the GLS model. The upper panels are for C-rich Miras, and the lower panels are for O-rich Miras.} \label{fig:accuracy}
\end{figure*}

The estimated frequency is considered as correct if $\Delta f=|\hat{f}-f_0|<C_f$ for each light curve. The estimation accuracies for the two methods are summarized in Table~\ref{tbl:accuracy} for several different values of $C_f$. We choose $C_f = 2.7\times 10^{-4}$ to stringently bind the one-to-one strip in Fig.~\ref{fig:fvsf}. Overall, the SP correctly estimates the period for 69.4\% of the light curves, while the LS model has an accuracy of 63.6\%. The improvement of SP over LS is more evident for C-rich Miras, with about 10\% higher accuracy, while the improvement for O-rich Miras is smaller, with about 3\% higher accuracy. The difference in performance arises because C-rich Miras often exhibit larger stochastic deviations that can be better captured by the SP model, while O-rich Miras have more stable light curves that can be modeled reasonably well with the LS method. We also compute the estimation accuracy of each method by grouping the light curves according to the number of observations, as shown in the left panels of Fig.~\ref{fig:accuracy}. The top and bottom rows show results for C- and O-rich Miras, respectively. The performance difference is once again more evident in the C-rich category. 

\begin{figure*}[hbtp]
\centering
\includegraphics[width = \textwidth]{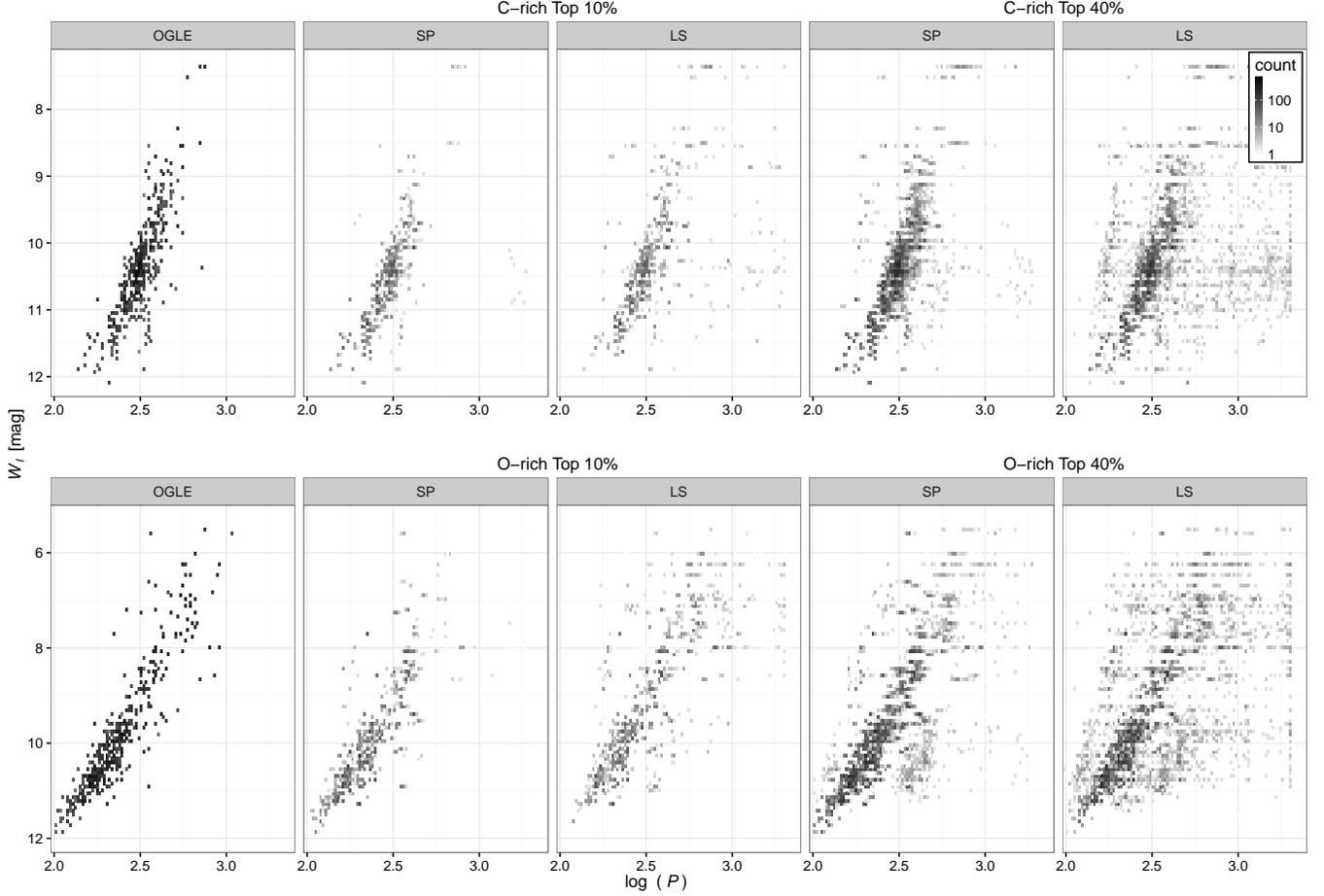}
\caption{Actual and reconstructed Period-Luminosity relations for C-rich (top) and O-rich (bottom) Miras. The leftmost column shows the actual PLRs using periods and $W_I$ magnitudes from \citet{Soszynski2009}. The other columns use the same magnitudes but periods based on the SP or LS algorithms, as indicated in each panel. Recovered PLRs are plotted for various subsets selected according to the {\tt conf} values obtained with the respective method.}\label{fig:pl.compare}
\end{figure*}

Note that accuracy is not a monotonic function of the number of observations, implying this is not a good indicator {\it per se} of the information content of the light curves for frequency (period) estimation. Thus, we define another metric, called {\it phase coverage}. Recall that the times of observation for a given light curve are $t_1,t_2,\cdots,t_n$. Given a period of $P$, these are converted into corresponding phases by $s_i = (t_i\ \mathrm{ mod }\ P)/P,\ i=1,2,\cdots, n\,$ in the closed interval $[0,1]$. Now, define $$J = \Big(\bigcup_i (s_i-l,s_i+l)\Big)\cap [0,1]\, ,$$ for a specific $l>0$, the phase coverage can be measured by $\lambda(J)$ where $\lambda(\cdot)$ is the Lebesgue measure (we choose $l=0.02$). $\lambda(J)$ describes the ``length'' of the union of the intervals $J$. For example, $\lambda(J) = 0.1$ for $J = (0.1,0.2)$, and $\lambda(J) = 0.2$ for $J = (0.1,0.2)\cup (0.5, 0.6)$.

We divide the light curves into 100 groups such that their $\lambda(J)$ is in one of the intervals $(k/100,(k+1)/100]$ for $k=0,1,\cdots,99$ and compute the estimation accuracy for each subset. The results for the two models are plotted in the middle column of Fig.~\ref{fig:accuracy}. Now the estimation accuracy is monotonically increasing as a function of phase coverage. The accuracy improvement of our method is highest when the phase coverage is around 0.5 for C-rich Miras. As the phase coverage approaches the extremes (0 or 1), the performance difference between the two methods diminishes.  At $\lambda(J) \approx 0$, both methods will fail because this is a hopeless situation. At the other extreme, when $\lambda(J)\approx 1$ and abundant information is available for frequency estimation, both methods have an accuracy close to 1.

The periodogram $S_{\rm SP}(f)$ of our model defined in Eqn.~\ref{eqn:periodogram} provides more information than just the optimal frequency. Suppose $f_1$ is the largest local maximal (global maximum) of $S_{\rm SP}(f)$, and $f_2$ is the second largest local maximal of $S_{\rm SP}(f)$. Now define {\tt conf}\ $=S_{\rm SP}(f_1)-S_{\rm SP}(f_2)\ge 0$. The value of {\tt conf} serves as a confidence measurement of the global optimal estimate in Eqn.~\ref{eqn:fhat}. Larger values of {\tt conf} indicate smaller uncertainty in our estimate, and thereby the estimate is more reliable. Now, let $c_0$ be the smallest value, and let $c_1, \dots, c_{100}$ be the $1$st--$100$th  percentiles of all the {\tt conf} values computed for all the light curves. Each light curve can be assigned to a percentile group if its {\tt conf} is in $(c_{k-1},c_{k}]$ for some $k\in\{1,2,\cdots,100\}$. After assigning all light curves by {\tt conf} to their corresponding percentile groups, the estimation accuracy in each group can be computed. The same procedure is applied to the GLS model, with the $p$-value of the F-statistics given in Eqn.~\ref{eq:glsP} for the top peak being used as its {\tt conf} measurement. The result is plotted in the right column of Fig.~\ref{fig:accuracy}. The accuracy of our SP method is much higher than the LS model in the top 40 groups. In particular, the accuracy of our method is higher than 90\% in the top 20 groups for both C- and O-rich Miras. 

\begin{figure}[h]
\centering
\includegraphics[width = 0.49\textwidth]{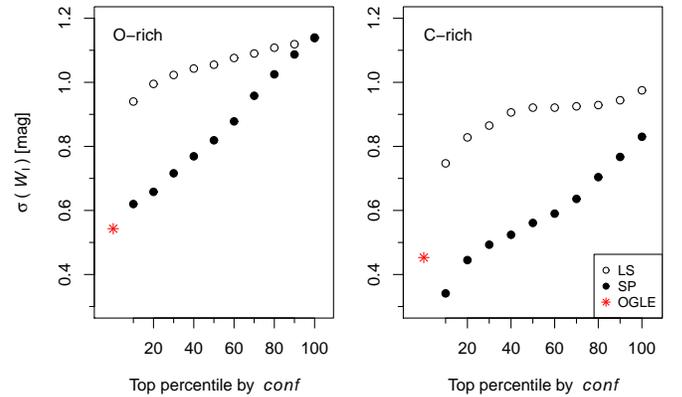}
\caption{Dispersion of reconstructed Wesenheit PLRs for different sets of artificial light curves, based on their {\tt conf} value. Top: O-rich PLRs. Bottom: C-rich PLRs. The starred symbols show the dispersion of the actual OGLE periods and $W_I$ magnitudes. The filled symbols show the dispersion of the recovered PLRs based on SP-derived periods. The open symbols show the corresponding values for LS-derived periods.}\label{fig:pl.sig}
\end{figure}

Light curves with high values of {\tt conf} are particularly reliable for constructing Period-Luminosity relations (hereafter, PLRs) based on the ``Wesenheit'' function \citep{Madore1982}. This function enables a simultaneous correction for the effects of dust attenuation and finite width of the instability strip by defining a new magnitude $W_I\!=\!I\!-\!1.55(V\!-\!I)$, where $V$ and $I$ are the mean magnitudes in those filters. Figure~\ref{fig:pl.compare} compares PLRs based on $W_I$ magnitudes and periods determined by OGLE and estimated with each of the two models. The top and bottom rows display the PLRs for C- and 

\begin{deluxetable}{ccrrr}
\renewcommand{\tabcolsep}{9pt}
\tablecaption{Comparison of estimation accuracies (\%)\label{tbl:accuracy}}
\tablewidth{0.49\textwidth}
\tablehead{\colhead{$C_f$} & \multirow{2}{*}{Method} & \multicolumn{3}{c}{Class}\\
\colhead{$(10^{-4})$} &   & \colhead{C-rich} & \colhead {O-rich} & \colhead{Both}}
\startdata
\multirow{2}{*}{1.0} & SP & 58.1 & 55.3 & 56.5 \\
                     & LS & 49.4 & 51.6 & 50.6 \\
\hline
\multirow{2}{*}{2.0} & SP & 69.6 & 63.8 & 66.3 \\
                     & LS & 60.1 & 60.6 & 60.4 \\
\hline
\multirow{2}{*}{2.7} & SP & 73.5 & 66.2 & 69.4 \\
                     & LS & 63.7 & 63.5 & 63.6 \\
\hline
\multicolumn{2}{c}{$N$ light curves} & 43,116 & 56,884 & \mcr{$10^5$}
\enddata
\tablecomments{Comparison of estimation accuracy for the SP and\\the GLS models. The accuracy is computed for several\\values of $C_f$ and estimated separately for C- and O-rich\\Miras; overall values are also given. The number of simu-\\lated light curves for each class is also listed.}
\end{deluxetable}
\vspace*{-30pt}

\noindent{O-rich Miras, respectively. The leftmost column shows the PLRs based on the actual OGLE periods, while the next two sets of columns show the corresponding relations based on SP or LS periods for the simulated light curves with the top 10\% and 40\% values of {\tt conf}.}

In order to provide a quantitative comparison of the improvement obtained with our SP method, we calculated the dispersion of the actual $W_I$ PLRs and their recovered counterparts as a function of {\tt conf} value as follows, separately for C- and O-rich Miras. First, we selected all objects of a given class with $2\!<\!\log P\!<\!3$. If the \citet{Soszynski2009} catalog did not provide a $V$ measurement for a given variable, the missing value was estimated through linear interpolation of the $(I,V\!-\!I)$ relation for objects of the same class within $|\Delta\log P|<0.05$~dex. We fitted a quadratic PLR $$m=a+b(\log P-2.3)+c(\log P-2.3)^2$$ with iterative $3\sigma$ clipping (removing $\sim 5$\% of the data). We then computed the dispersion of the initially selected OGLE sample about the best-fit relation, including outliers. This yielded ``benchmark'' dispersions of 0.45 \& 0.54~mag for C- \& O-rich variables, respectively. Keeping the best-fit relation fixed, we computed the dispersion of recovered PLRs using all artificial light curves within a certain range of {\tt conf} (top 10\%, top 20\%, $\dots$), using the periods and {\tt conf} values derived by the SP or the LS method. As in the case of the OGLE samples, we only considered objects with $2\!<\!\log P\!<\!3$. The results are plotted in Fig.~\ref{fig:pl.sig}. The SP subsamples exhibit lower

\begin{deluxetable}{llrrrr}
\renewcommand{\tabcolsep}{9pt}
\tablecaption{OGLE LMC Miras from \citet{Soszynski2009}\label{tbl:oglemira}}
\tablewidth{0.49\textwidth}
\tablehead{\colhead{OGID} & \colhead{T} & \colhead{$\bar{I}$} & \colhead{$\bar{V}$} & \colhead{P} & \colhead {fl}\\
\colhead{} & \colhead{} & \colhead{(mag)} & \colhead{(mag)} & \colhead{(d)} & \colhead{}}
\startdata
00082 & O & 14.241 & 16.509 & 164.84 & \\
00094 & C & 15.120 & 18.885 & 332.30 & \\
00098 & C & 15.159 & 17.921 & 323.10 & \\
00115 & C & 14.932 & 16.947 & 176.13 & \\
00355 & O & 14.199 & 16.219 & \mcr{154.59} & \\
\enddata
\tablecomments{Objects with missing data and extrapolated mean\\$V$ magnitudes are identified with a * in the flag column.\\(This table is available in its entirety in machine-readable\\form.)}
\end{deluxetable}
\vspace*{-30pt}

\noindent{(or at worst, equal) dispersions than their LS counterparts for all percentiles and for both subtypes. As discussed previously, the improvement provided by our method is strongest for C-rich Miras and diminishes in significance as one includes light curves with progressively lower confidence values.}

\section{Summary} \label{sec.discussion}
In this paper, we developed a nonlinear SP Gaussian process model for estimating the periods of sparsely sampled quasi-periodic light curves, motivated by the desire to detect Miras in an existing set of observations of M33. We conducted a large-scale high-fidelity simulation of Mira light curves as observed by the DIRECT/M33SSS surveys to compare our model with the GLS method. Our model shows improved accuracy under various metrics. The simulation data set is provided as a testbed for future comparison with other methods. The SP model will be used in a companion paper to search for Miras in M33, estimate their periods, and study the resulting PLRs.

\ \par
SH was partially supported by Texas A\&M University-NSFC Joint Research Program. WY \& LMM acknowledge financial support from the NSF through AST grant \#1211603 and from the Mitchell Institute for Fundamental Physics and Astronomy at Texas A\&M University. JZH was partially supported by NSF grant DMS-1208952. The authors acknowledge the Texas A\&M University Brazos HPC cluster that contributed to the research reported here.

\bibliographystyle{apj}
\bibliography{m33gp}

\begin{appendix}
\noindent{{\it Simulated light curves:} A tarfile, containing $10^5$ simulated light curves are generated following the procedure of \S\ref{sec.construct.test}. Each light curve is stored in one file with three columns: MJD, $I$ magnitude, and uncertainty. The file name, e.g., lc006788.dat is generated sequentially and is only meant for bookkeeping purposes. A mapping between simulated light curve ID and the original OGLE object is given in the file ``lc.dat'', which can also be found in the tarfile.}

\noindent{{\it Mira variables:} Table\ref{tbl:oglemira} summarizes the relevant properties of OGLE LMC Miras from \citet{Soszynski2009} that were used to simulate the light curves: OGLE ID, main period and mean $I$ \& $V$ magnitudes. It includes some extrapolated values of $V$ for objects with missing data (suitably identified with a ``*''). This table can be used to compare true versus derived periods and to generate Period-Luminosity relations.}

\noindent{{\it Software:} The related software package, \texttt{varStar}, has been released under a GPL3 license \citep{He2016}. The active software development repository can be found at \url{github.com/shiyuanhe/varStar}.}

\center{\sc Pseudo-code}
\begin{algorithmic}
\algrenewcommand\algorithmicindent{0.5em}
\Procedure{quasinewton}
\textit{\ \it Quasi-Newton's Method with Grid Search}\\
\textbf{Input:} Maximal and minimal trial frequencies $f_M\!>\!f_m\!>\!0$; frequency step $\Delta f$; $n$ observations $\{t_i,y_i,\sigma_i\}$.\\ 
\textbf{Output:} Periodogram $S(f)$ evaluated at the trial frequencies.\\
\begin{algorithmic}[1]
  \State Initialize $\vtheta^{(0)}$ and $\mathbf{H}^{(0)}$, and $f\gets f_m$;

\For{$f \in \{f_m, f_m + \Delta f, f_m + 2\Delta f, \cdots, f_M\}$}

  \State $p \gets 0$;

  \Repeat
  \State $\mathbf{t}_p \gets -\mathbf{H}^{(p)} \frac{\partial}{\partial \vtheta}
Q(\vtheta^{(p)}, f)$;

 \State $\vtheta^{(p+1)}\gets \vtheta^{(p)} +\alpha_p \mathbf{t}_p$ and
 the step size $\alpha_p$ satisfying the Wolfe condition;

\State $\mathbf{d}_p \gets \vtheta^{(p+1)}-\vtheta^{(p)}$,
$\mathbf{e}_p \gets \frac{\partial}{\partial \vtheta} Q(\vtheta^{(p+1)}, f)- 
\frac{\partial}{\partial \vtheta} Q(\vtheta^{(p)}, f)$;

\State $\rho_p= 1/\mathbf{d}_p^T\mathbf{e}_p$;

\State $\mathbf{H}^{(p+1)}\gets (\mathbf{I}-\rho_p
\mathbf{d}_p\mathbf{e}_p^T)
\mathbf{H}^{(p)} (\mathbf{I}-\rho_p \mathbf{e}_p\mathbf{d}_p^T) 
+\rho_p\mathbf{d}_p\mathbf{d}_p^T$;\Comment{BFGS update}

  \State $p \gets p + 1$;
  \Until{$\Vert \mathbf{d}_{p-1}\Vert < \epsilon$}

  \State $\widehat{\vtheta}_{f} \gets \vtheta^{(p)}$, and 
$S(f)\gets Q(\widehat{\vtheta}_{f},f)$;

 \State $\mathbf{H}^{(0)} \gets \mathbf{H}^{(p)}$ and
 $\vtheta^{(0)}\gets \vtheta^{(p)}$; 
\Comment{Save warm start value for next trial frequency}
\EndFor
\end{algorithmic}
\end{algorithmic}
\end{appendix}
\end{document}